\documentclass[a4paper,11pt]{JHEP3}
\usepackage[OT4]{fontenc}
\usepackage{amsmath}
\usepackage[active]{srcltx}

\newcommand{\chiral}[3]{\!\big[\hspace*{-2pt}\begin{array}{c} \\[-19pt]{\scriptscriptstyle #1}\\[-9pt]{\scriptscriptstyle #2 #3}\end{array}\hspace*{-2pt}\big]}

\def\podkr{\hspace*{-1pt}{\begin{array}{c}\\[-18pt]\underline{\hspace*{5pt}}\end{array}}\hspace*{-1pt}}
\def\upodkr{\hspace*{-1pt}{\begin{array}{c}\\[-20pt]\underline{\hspace*{5pt}}\end{array}}\hspace*{-1pt}}

\def\bra#1{\left\langle#1\right|}
\def\ket#1{\left|#1\right\rangle}

\newcommand{\RRe}{{\scriptscriptstyle \rm RR, e}}
\newcommand{\RRo}{{\scriptscriptstyle \rm RR, o}}
\newcommand{\NN}{{\scriptscriptstyle \rm NN}}

\author{
     Leszek Hadasz\footnote{e-mail: hadasz@th.if.uj.edu.pl} \\
	 M. Smoluchowski Institute of Physics,
	 Jagiellonian University,\\
	 W.~Reymonta 4,
	 30-059~Krak\'ow, Poland

}
\author{
     Zbigniew Jask\'{o}lski \footnote{e-mail: jask@ift.uni.wroc.pl}\\
	 Institute of Theoretical Physics,
	 University of Wroc{\l}aw,\\
	 pl. M. Borna 1, 95-204~Wroc{\l}aw, Poland
}
\author{
     Paulina Suchanek \footnote{e-mail: paulina@ift.uni.wroc.pl}\\
     DESY Theory Group, Hamburg, Germany,\\
     and\\
	 Institute of Theoretical Physics,
	 University of Wroc{\l}aw,\\
	 pl. M. Borna 1, 95-204~Wroc{\l}aw, Poland
}

\abstract{General  1-point toric blocks in all sectors of
N=1 superconformal field theories are analyzed. The
recurrence relations for blocks coefficients are derived by calculating their residues and large $\Delta$ asymptotics.
\vspace*{1cm}}

\title{ Recurrence relations for toric N=1 superconformal blocks }

\preprint{}

\begin{document}

\section{Introduction}

Correlation functions in any 2-dimensional
CFT can be expressed in terms of three-point coupling
constants and some universal, model independent functions called conformal blocks  \cite{Belavin:1984vu}.
Such decompositions are not unique and the
equivalence of various representations
yields strong restrictions for
coupling constants.
Simplest conditions of this kind  ---
the crossing symmetry of the 4-point function on the sphere
and the modular invariance of the 1-point function on the torus --- turned out to be
 sufficient for the consistency of all multi-point amplitudes
on closed, oriented surfaces of arbitrary genus
\cite{Sonoda:1988fq}. From this point of view the 4-point spheric  and the 1-point toric
conformal blocks are of main interest in any CFT. Both objects are defined
as power series of corresponding modular parameters with coefficients expressed
in terms of the inverse of the Gram matrix of a Virasoro  algebra Verma module.
In this form a direct calculation of higher order terms
is prohibitively complicated.

An efficient recursive technique  of calculating
 4-point conformal blocks on the sphere was developed long time ago by Al.~Zamolodchikov
\cite{Zamolodchikov:ie,Zamolodchikov:2,Zamolodchikov:3}.
This method was used for numerical tests of crossing symmetry
 in the Liouville field theory \cite{Zamolodchikov:1995aa} and
in  the $c\to 1$ limit of minimal models
\cite{Runkel:2001ng}. It was also applied for the numerical analysis of
the classical limit of the conformal blocks and
for the verification of new relations in the classical geometry of hyperbolic surfaces
\cite{Hadasz:2005gk}.
Recently in the context of the AGT relation Zamolodchikov's method
has been extended to the 1-point toric blocks \cite{Poghossian:2009mk,Hadasz:2009db}.
It found its application  in one of the first proofs of the AGT correspondence \cite{Fateev:2009aw}.
It was also used in  \cite{Hadasz:2009db} to prove the relations between 1-point toric and 4-point spheric
conformal blocks conjectured by Poghossian
\cite{Poghossian:2009mk}.

A recursion representation of 4-point spheric blocks in the N=1 superconformal
field theories
 was first derived
in the Neveu-Schwarz sector
\cite{Hadasz:2006sb,Belavin:2007zz,Belavin:2007gz,Belavin:2007eq,Hadasz:2007nt}.
The extension to the Ramond sector initiated in \cite{Hadasz:2008dt}
was recently completed in \cite{Suchanek:2010kq}. These results clarified the structure of the N=1 superconformal blocks
and paved the way for investigations of their analytical properties \cite{Hadasz:2007wi,Chorazkiewicz:2008es,Chorazkiewicz:2011zd}.
They were also used for numerical verifications of crossing symmetry in the N=1 superconformal
Liouville field theory \cite{Belavin:2007gz,Belavin:2007eq,Suchanek:2010kq}.

In the present paper we address the problem of recursion representation of the 1-point toric
conformal blocks in the N=1 SCFT.
Our main motivation is the problem of modular invariance of 1-point functions on the torus
in the N=1 superconformal Liouville field theory with the structure constants
derived in \cite{Rashkov:1996jx,Poghosian:1996dw}.
The corresponding  problem
in the Liouville theory  was solved
by showing that the modular invariance of a generic 1-point function on the torus
is equivalent to the crossing symmetry of a special 4-point function on the sphere \cite{Hadasz:2009sw}.
An essential step of this reasoning is a relation between the modular and
the fusion matrices which can be derived using Poghosian identities \cite{Poghossian:2009mk,Hadasz:2009db}.
One may expect a similar, although  more complicated
mechanism in the N=1 superconformal case.
The recurrence representations which in the bosonic case were basic tools in analyzing relations between toric and spheric N=1 blocks
are first steps along this line.
They are also of some interest for the recently discovered extension of the AGT relation where
on the CFT side the N=1 superconformal Liouville field theory shows up
\cite{Belavin:2011pp,Bonelli:2011jx,Belavin:2011tb,Bonelli:2011kv,Ito:2011mw,Belavin:2011sw}.

The paper is organized as follows. Section 2 contains a detailed  discussion of  toric blocks in all sectors of
N=1 superconformal field theories. In Section 3 we calculate the residues of blocks coefficients. The method employed is
a simple extension of the techniques developed for the spheric case. The main technical point is discussed in Section 4 where
we calculate the large $\Delta$ asymptotics. The derivation is based on properties of the Gram matrix and the matrix elements of chiral
vertex operators. Proofs of these properties  are given in Appendices A and B.
In Section 5 we derive the recursion relations for N=1 superconformal toric blocks which are the main results of the present paper.
Explicit formulae for first few block's coefficients are listed in Appendix C.

\section{1-point toric blocks in N=1 superconformal field theories}

In the N=1 superconformal field theory on a torus the basic independent 1-point  functions
are those  of  super-primary NS fields
$\phi_{\lambda,\bar\lambda}(z,\bar z)$ and their even primary descendants
\[
\widetilde \phi_{\lambda,\bar\lambda}(z,\bar z)= \left\{S_{-1/2},\left[\bar S_{-1/2},\phi_{\lambda,\bar\lambda}(z,\bar z)\right]
\right\}.
\]
In our notation $\lambda,\bar\lambda$ parameterize the left and the right conformal weights of $\phi_{\lambda,\bar\lambda}(z,\bar z)$
$$
\Delta_\lambda  =\frac{Q^2}{8} - \frac{\lambda^2}{8},\hskip 1cm \Delta_{\bar\lambda}  = \frac{Q^2}{8} - \frac{\bar\lambda^2}{8},
$$
and $Q = b+ b^{-1}$ is related to the central charge $c$ of the NSR algebra by $c = \frac32 + 3Q^2.$

\subsection{${\rm \bf NS}$ and $ \widetilde{\rm \bf NS}$   sectors}

The 1-point function of $\phi_{\lambda,\bar\lambda}$ on a torus
with the modular parameter $\tau$
can be written as
\begin{eqnarray}
\label{coorfunct:1}
\langle \phi_{\lambda,\bar \lambda} \rangle_{\scriptscriptstyle \rm NS }
&=& (q \bar q)^{-\frac{c}{24}}
\sum\limits_{(\Delta,\bar\Delta)}\; \sum\limits_{f,\, \bar f\in \frac12\mathbb{N}\cup\{0\}} \;
 q^{\Delta + f} \, \bar{q}^{\bar \Delta + \bar f}   \\
 \nonumber
 &&\hspace{-40pt}\times  \hspace{-35pt}
 \sum\limits_{\begin{array}{c}\scriptstyle {f=|M|+|K|=|N|+|L|}\\[-6pt]
 \scriptstyle {\bar f=|\bar M|+|\bar K|=|\bar N|+|\bar L|}
\end{array}} \hspace{-30pt}
\Big[ B^f_{\Delta} \Big]^{MK,NL}
\Big[B^{\bar f}_{\bar \Delta} \Big]^{\bar M \bar K, \bar N \bar L}
\bra{\nu_{\Delta,MK} \otimes {\nu}_{\bar \Delta,\bar M\bar K}}  \phi_{\lambda,\bar \lambda}(1,1)
| \nu_{\Delta,NL} \otimes{\nu}_{\bar\Delta,\bar N \bar L}  \rangle
\end{eqnarray}
where $
q = e^{2 \pi i \tau}
$ and the sum runs over the whole spectrum of the NS theory.
The matrices
$\Big[ B^f_{\Delta} \Big]^{MK,NL}$, $\Big[B^{\bar f}_{\bar \Delta} \Big]^{\bar M \bar K, \bar N \bar L}$
are inverse to the Gram matrices
$$
\Big[ B^f_{\Delta} \Big]_{MK,NL}=
\bra{\nu_{\Delta,MK} }
 \nu_{\Delta,NL}   \rangle,
 \;\;\;\;\;\;
 \Big[B^{\bar f}_{\bar \Delta} \Big]_{\bar M \bar K, \bar N \bar L}
 =
 \bra{{\nu}_{\bar \Delta,\bar M\bar K}}
 {\nu}_{\bar\Delta,\bar N \bar L}  \rangle.
$$
calculated in the standard bases in the corresponding NS Verma modules:
\begin{eqnarray*}
\nu_{\Delta,MK}
&=&
L_{-M}S_{-K} \nu_{\Delta}
\; = \;
L_{-m_j}\ldots L_{-m_1}S_{-k_i}\ldots S_{-k_1}\nu_{\Delta}\,,
\\
&&
k_i > \ldots  > k_1 ,
\hspace{10pt} k_s\in \mathbb{N}-\textstyle \frac12, \hspace{20pt}
m_j \geqslant \ldots \geqslant  m_1,\hspace{10pt} m_r\in \mathbb{N}
\\
L_0\nu_{\Delta}&=&\Delta\nu_{\Delta},
\hskip 5mm
(-1)^F\nu_{\Delta}=\nu_{\Delta},
\hskip 5mm
L_m\nu_{\Delta}= S_k\nu_{\Delta} =0.
\end{eqnarray*}
For $K,L$ and $\bar K,\bar L$ of the same parity one has
\begin{eqnarray}
\label{three:point:function}
\nonumber
 \bra{\nu_{\Delta,MK} \otimes {\nu}_{\bar \Delta,\bar M\bar K}}  \phi_{\lambda,\bar \lambda}(1,1)
| \nu_{\Delta,NL} \otimes{\nu}_{\bar\Delta,\bar N \bar L}  \rangle
 & = &
\\ \nonumber
&&\hspace{-130pt}
  \rho _{\scriptscriptstyle \rm NN }(\nu_{\Delta,MK}, \nu_{\lambda} , \nu_{\Delta,NL}) \,
  \rho_{\scriptscriptstyle \rm NN }( \nu_{\bar\Delta,\bar M \bar K},   \nu_{\bar\lambda} ,  \nu_{\bar\Delta,\bar N \bar K})\,
  C^{\lambda,\bar\lambda}_{\Delta,\bar\Delta},
  \\[-7pt]
  \\[-7pt] \nonumber
\bra{\nu_{\Delta,MK} \otimes {\nu}_{\bar \Delta,\bar M\bar K}}
\tilde \phi_{\lambda,\bar \lambda}(1,1)
| \nu_{\Delta,NL} \otimes{\nu}_{\bar\Delta,\bar N \bar L}  \rangle
 & = &
 \\ \nonumber
&&\hspace{-130pt}
  \rho^*_{\scriptscriptstyle \rm NN }(\nu_{\Delta,MK}, \ast\nu_{\lambda} , \nu_{\Delta,NL}) \,
  \rho^*_{\scriptscriptstyle \rm NN }( \nu_{\bar\Delta,\bar M \bar K},   \ast \nu_{\bar\lambda} ,  \nu_{\bar\Delta,\bar N \bar K})\,
 \tilde C^{\lambda,\bar\lambda}_{\Delta,\bar\Delta},
\end{eqnarray}
where $C^{\lambda,\bar\lambda}_{\Delta,\bar\Delta}$, $\tilde C^{\lambda,\bar\lambda}_{\Delta,\bar\Delta}$
 are the three-point constants
\[
C^{\lambda,\bar\lambda}_{\Delta,\bar\Delta}
=
\langle \nu_{\Delta} \otimes {\nu}_{\bar \Delta}|  \phi_{\lambda,\bar \lambda}(1,1)| \nu_{\Delta} \otimes{\nu}_{\bar\Delta,}  \rangle,
\hskip 5mm
\tilde C^{\lambda,\bar\lambda}_{\Delta,\bar\Delta}
=
\langle \nu_{\Delta} \otimes {\nu}_{\bar \Delta}|  \tilde \phi_{\lambda,\bar \lambda}(1,1)| \nu_{\Delta} \otimes{\nu}_{\bar\Delta,}  \rangle,
\]
and $\rho_{\scriptscriptstyle \rm NN }$, $\rho^{*}_{\scriptscriptstyle \rm NN }$ are 3-point conformal blocks in the NS sector.
In the formulae above and in the rest of the paper we follow
the notation conventions of \cite{Chorazkiewicz:2011zd}.

The toric conformal blocks are defined by
 \begin{eqnarray}
  \nonumber
\mathcal{F}_{ \Delta}^{\upodkr\lambda}(q)
    &=& q^{\Delta -\frac{c}{24}} \,
\sum_{f\in \frac12\mathbb{N}} q^{f} \,
    F^{\upodkr\lambda, f}_{ \Delta},
\\[-10pt]
\label{blockCoeff}
\\[-7pt]
\nonumber
F^{\upodkr\lambda, f}_{ \Delta} &=&
\sum_{M,K,N,L}
 \rho^{\podkr}_{\scriptscriptstyle \rm NN }\,(\nu_{\Delta,MK}, \underline{\hspace*{5pt}}\,\nu_{\lambda} , \nu_{\Delta,NL})
\left[ B^f_{\Delta} \right]^{MK,NL},
\end{eqnarray}
\vspace*{-15pt}

\noindent
where the symbol $\podkr$ stands for the star or the lack of it.
The  1-point functions can then be represented as
\begin{eqnarray*}
 \langle \phi_{\lambda,\bar \lambda} \rangle_{\scriptscriptstyle \rm NS }
 &=&
  \sum_{(\Delta,\bar\Delta)}
 \mathcal{F}_{ \Delta}^{\lambda}(q)\,  \mathcal{F}_{ \bar \Delta}^{\bar\lambda}(\bar q)
\,
    C^{\lambda,\bar\lambda}_{\Delta,\bar\Delta},
\\
 \langle \tilde \phi_{\lambda,\bar \lambda} \rangle_{\scriptscriptstyle \rm NS }
 &=&
  \sum_{(\Delta,\bar\Delta)}
 \mathcal{F}_{ \Delta}^{\ast \lambda}(q)\,   \mathcal{F}_{ \bar \Delta}^{\ast\bar\lambda}(\bar q)
\,
  \tilde C^{\lambda,\bar\lambda}_{\Delta,\bar\Delta}.
\end{eqnarray*}
In $ \widetilde{\rm \bf NS}$
sector one introduces the modified conformal blocks \cite{Matsuo:1986vc}
\begin{eqnarray}
\label{tildeblocks}
\widetilde{\cal F}^{\underline{\hspace*{3pt}}\lambda}_{\Delta}(q)
& = &
q^{\Delta -\frac{c}{24}} \, \sum_{f} (-1)^{2f}q^{f} \, F^{\underline{\hspace{4pt}}\,\lambda, f}_{ \Delta}
\end{eqnarray}
and
the 1-point functions take the form
\begin{eqnarray*}
 \langle \phi_{\lambda,\bar \lambda} \rangle_{\widetilde{\scriptscriptstyle \rm NS }}
 &=&
  \sum_{(\Delta,\bar\Delta)}
\widetilde{ \mathcal{F}}_{ \Delta}^{\lambda}(q)\,
\widetilde{ \mathcal{F}}_{ \bar \Delta}^{\bar\lambda}(\bar q)
\,
    C^{\lambda,\bar\lambda}_{\Delta,\bar\Delta},
\\
 \langle \tilde \phi_{\lambda,\bar \lambda} \rangle_{\widetilde{\scriptscriptstyle \rm NS }}
 &=&
  \sum_{(\Delta,\bar\Delta)}
\widetilde{ \mathcal{F}}_{ \Delta}^{\ast \lambda}(q)\,
\widetilde{ \mathcal{F}}_{ \bar \Delta}^{\ast\bar\lambda}(\bar q)
\,
  \tilde C^{\lambda,\bar\lambda}_{\Delta,\bar\Delta}.
\end{eqnarray*}

\subsection{ ${\rm \bf R}$ and ${\rm \bf \widetilde R}$ sectors}

The toric 1-point functions in the $R$ sector read
\begin{eqnarray}
\label{1-point R}
\langle \phi_{\lambda,\bar \lambda} \rangle_{\scriptscriptstyle \rm R}
 &=&  (q \bar q)^{-\frac{c}{24}}
\sum\limits_{(\beta,\bar\beta)}\; \sum\limits_{f,\,\bar f=0}^\infty \;
 q^{\Delta_\beta + f} \, \bar{q}^{\Delta_{\bar\beta} + \bar f}
\\\nonumber
&\times&
\hspace{-33pt}
\sum\limits_{\begin{array}{c}\scriptstyle {f=|M|+|K|=|N|+|L|}\\[-6pt]
 \scriptstyle {\#  K+\#  L\in 2\mathbb{N}}
\end{array}}
 \hspace{-10pt}
\Big[ B^f_{c\,\beta} \Big]^{MK,NL}
\hspace{-15pt}
\sum\limits_{\begin{array}{c} \scriptstyle {\bar f=|\bar M|+|\bar K|=|\bar N|+|\bar L|}\\[-6pt]
 \scriptstyle {\# \bar K+\#\bar  L\in 2\mathbb{N}}
\end{array}} \hspace{-10pt}
\Big[B^{\bar f}_{c\,\bar \beta} \Big]^{\bar M \bar K, \bar N \bar L}
\\
\nonumber
&\times &
\bra{L_{-M}S_{-K} \bar L_{-\bar M} \bar S_{- \bar K} w^{+}_{\beta, \bar \beta}}
\phi_{\lambda,\bar \lambda}(1,1)
\ket{ L_{- N} S_{-L}\bar L_{-\bar N} \bar S_{- \bar L} w^{+}_{\beta, \bar \beta}}.
\end{eqnarray}
The matrices
$\Big[ B^f_{c\,\beta} \Big]^{MK,NL}$, $\Big[B^{\bar f}_{c\,\bar \beta} \Big]^{\bar M \bar K, \bar N \bar L}$
are inverse to the Gram matrices
$$
\Big[ B^f_{c\,\beta} \Big]_{MK,NL}=
\bra{{w}^+_{\beta,MK} }
 {w}^+_{\beta,NL}   \Big\rangle,
 \;\;\;\;\;\;
 \Big[B^{\bar f}_{c\,\bar \beta} \Big]_{\bar M \bar K, \bar N \bar L}
 =
 \bra{{w}^+_{\bar \beta,\bar M\bar K}}
 {w}^+_{\bar\beta,\bar N \bar L}  \Big\rangle,
$$
calculated in the standard bases in the corresponding R Verma modules
${\cal W}_{\beta}$, ${\cal W}_{\bar\beta}$:
\begin{eqnarray*}
{w}^+_{\beta,MK}
&=&
L_{-M}S_{-K} {w}^+_\beta
\; = \;
L_{-m_j}\ldots L_{-m_1}S_{-k_i}\ldots S_{-k_1}{w}^+_\beta\,,
\\
&&
k_i > \ldots  > k_1 ,
\hspace{10pt} k_s\in \mathbb{N}\cup\{0\}, \hspace{20pt}
m_j \geqslant \ldots \geqslant  m_1,\hspace{10pt} m_r\in \mathbb{N},
\\
L_0{w}^+_\beta&=&\Delta_\beta{w}^+_\beta,\;\;\;\Delta_\beta=\textstyle \frac{c}{24}-\beta^2,
\\
S_0  w^\pm_{\beta} &=&i {\rm e}^{\mp i\frac{\pi}{4}} \beta  w^\mp_\beta\neq 0,\;\;\;
(-1)^F{w}^+_\beta={w}^+_\beta,
\\
L_m{w}^+_\beta&=& S_k{w}^+_\beta=0\;\;\;{\rm for}\;\;\;m,k>0,
\end{eqnarray*}
and
$$
w^{+}_{\beta, \bar \beta}=
 \frac{1}{\sqrt{2}}\left( w^+_\beta \otimes  w^+_{\bar\beta} -i\,
 w^-_\beta \otimes  w^-_{\bar\beta}\right).
$$
The chiral decompositions of $\phi_{\lambda,\bar \lambda}$
and $\tilde \phi_{\lambda,\bar \lambda}$
in the Ramond sector
take the form
\cite{Suchanek:2010kq}
\begin{equation}
\label{phi_RR}
\begin{array}{lll}
\phi_{\lambda,\bar \lambda}&=&
C^{\lambda\,\bar \lambda(+)}_{\beta, \bar \beta}
  \left(
  V_{\scriptscriptstyle \rm \bf e}^+\chiral{\Delta_\lambda}{\beta\:}{\:\beta}
   \otimes
  V_{\scriptscriptstyle \rm \bf e}^+\chiral{\Delta_{\bar\lambda}}{\bar\beta\:}{\:\bar\beta}
            - i
  V_{\scriptscriptstyle \rm \bf o}^+\chiral{\Delta_\lambda}{\beta\:}{\:\beta}
   \otimes
  V_{\scriptscriptstyle \rm \bf o}^+\chiral{\Delta_{\bar\lambda}}{\bar\beta\:}{\:\bar\beta}
\right)
\\
&+&
C^{\lambda\,\bar \lambda(-)}_{\beta, \bar \beta}
\left(
  V_{\scriptscriptstyle \rm \bf e}^-\chiral{\Delta_\lambda}{\beta\:}{\:\beta}
   \otimes
  V_{\scriptscriptstyle \rm \bf e}^-\chiral{\Delta_{\bar\lambda}}{\bar\beta\:}{\:\bar\beta}
            - i
  V_{\scriptscriptstyle \rm \bf o}^-\chiral{\Delta_\lambda}{\beta\:}{\:\beta}
   \otimes
  V_{\scriptscriptstyle \rm \bf o}^-\chiral{\Delta_{\bar\lambda}}{\bar\beta\:}{\:\bar\beta}
\right),
\\            [6pt]
\tilde\phi_{\lambda,\bar \lambda}&=&
C^{\lambda\,\bar \lambda(+)}_{\beta, \bar \beta}
  \left(
 i V_{\scriptscriptstyle \rm \bf e}^+\chiral{*\Delta_\lambda}{\beta\:}{\:\beta}
   \otimes
   V_{\scriptscriptstyle \rm \bf e}^+\chiral{*\Delta_{\bar\lambda}}{\bar\beta\:}{\:\bar\beta}
            +
   V_{\scriptscriptstyle \rm \bf o}^+\chiral{*\Delta_\lambda}{\beta\:}{\:\beta}
   \otimes
   V_{\scriptscriptstyle \rm \bf o}^+\chiral{*\Delta_{\bar\lambda}}{\bar\beta\:}{\:\bar\beta}
\right)
\\
&+&
C^{\lambda\,\bar \lambda(-)}_{\beta, \bar \beta}
\left(
 i V_{\scriptscriptstyle \rm \bf e}^-\chiral{*\Delta_\lambda}{\beta\:}{\:\beta}
   \otimes
   V_{\scriptscriptstyle \rm \bf e}^-\chiral{*\Delta_{\bar\lambda}}{\bar\beta\:}{\:\bar\beta}
            +
   V_{\scriptscriptstyle \rm \bf o}^-\chiral{*\Delta_\lambda}{\beta\:}{\:\beta}
   \otimes
   V_{\scriptscriptstyle \rm \bf o}^-\chiral{*\Delta_{\bar\lambda}}{\bar\beta\:}{\:\bar\beta}
\right),
\end{array}
\end{equation}
where the chiral vertex operators are defined in terms of 3-point blocks   \cite{Chorazkiewicz:2011zd}
\begin{equation}
\label{chiral:vertex:RNR}
\begin{array}{rcl}
 \bra{w^+_{\beta,MK}}
 V_{\scriptscriptstyle \rm \bf f}^\pm\chiral{\Delta_\lambda}{\beta\:}{\:\beta}
(z)
 \ket{w^+_{\beta,NL}}
&=& \rho^{(\pm)}_{\scriptscriptstyle \rm RR, f}(w^+_{\beta,MK},\nu_\lambda,w^+_{\beta,NL}|z),
\\[6pt]
 \bra{w^+_{\beta,MK}}
  V_{\scriptscriptstyle \rm \bf f}^\pm\chiral{*\Delta_\lambda}{\beta\:}{\beta\:}
(z)
 \ket{w^+_{\beta,NL}}
&=& \rho^{(\pm)}_{\scriptscriptstyle \rm RR, \bar{f}}(w^+_{\beta,MK},*\nu_\lambda,w^+_{\beta,NL}|z).
\end{array}
\end{equation}
For $ \# K+\, \# L\in  2\mathbb{N}$ one has in particular:
\begin{eqnarray}
\nonumber
&& \hspace{-80pt} \bra{L_{ -M} S_{-K}\bar L_{-\bar M} \bar S_{- \bar K} w^{+}_{\beta, \bar \beta}}
 \phi_{\lambda,\bar \lambda}(1,1)
 \ket{L_{-N}S_{-L} \bar L_{-\bar N}\bar S_{- \bar L} w^{+}_{\beta, \bar \beta}}
\\[4pt]
\label{Ramond:3ptcorrelator:1}
&=&
C^{\lambda\,\bar \lambda(+)}_{\beta, \bar \beta}
\rho^{(+)}_{\scriptscriptstyle \rm RR, e}(w^+_{\beta,MK}, \nu_\lambda,w^+_{\beta,NL}) \,
\rho^{(+)}_{\scriptscriptstyle \rm RR, e}(w^+_{\bar\beta,\bar M\bar K}, \nu_{\bar \lambda},w^+_{\bar\beta,\bar N\bar L})
\\[4pt]
\nonumber
& + &
C^{\lambda\,\bar \lambda(-)}_{\beta, \bar \beta}
\rho^{(-)}_{\scriptscriptstyle \rm RR, e}(w^+_{\beta,MK}, \nu_\lambda,w^+_{\beta,NL}) \,
\rho^{(-)}_{\scriptscriptstyle \rm RR, e}(w^+_{\bar\beta,\bar M\bar K}, \nu_{\bar \lambda},w^+_{\bar\beta,\bar N\bar L}),
\end{eqnarray}
and
\begin{eqnarray}
\nonumber
&& \hspace{-70pt} \bra{L_{ -M} S_{-K}\bar L_{-\bar M} \bar S_{- \bar K} w^{+}_{\beta, \bar \beta}}
 \tilde \phi_{\lambda,\bar \lambda}(1,1)
 \ket{L_{-N}S_{-L} \bar L_{-\bar N}\bar S_{- \bar L} w^{+}_{\beta, \bar \beta}}
\\[4pt]
\label{Ramond:3ptcorrelator:2}
&=&
iC^{\lambda\,\bar \lambda(+)}_{\beta, \bar \beta}
\rho^{(+)}_{\scriptscriptstyle \rm RR, o}(w^+_{\beta,MK}, *\nu_\lambda,w^+_{\beta,NL}) \,
\rho^{(+)}_{\scriptscriptstyle \rm RR, o}(w^+_{\bar\beta,\bar M\bar K}, *\nu_{\bar \lambda},w^+_{\bar\beta,\bar N\bar L})
\\[4pt]
\nonumber
&+&
iC^{\lambda\,\bar \lambda(-)}_{\beta, \bar \beta}
\rho^{(-)}_{\scriptscriptstyle \rm RR, o}(w^+_{\beta,MK}, *\nu_\lambda,w^+_{\beta,NL}) \,
\rho^{(-)}_{\scriptscriptstyle \rm RR, o}(w^+_{\bar\beta,\bar M\bar K}, *\nu_{\bar \lambda},w^+_{\bar\beta,\bar N\bar L}).
\end{eqnarray}
1-point function (\ref{1-point R}) can then be written as
\begin{eqnarray*}
\langle \phi_{\lambda,\bar \lambda} \rangle_{\scriptscriptstyle \rm R}
&=&
\sum_{(\beta,\bar\beta)}
C^{\lambda\,\bar \lambda(+)}_{\beta, \bar \beta}
 \mathcal{F}_{\beta}^{\, \lambda(+)}(q)\, {\mathcal{F}}_{ \bar \beta}^{\, \bar  \lambda(+)}(\bar q)
 +
 \sum_{(\beta,\bar\beta)}
C^{\lambda\,\bar \lambda(-)}_{\beta, \bar \beta}
 \mathcal{F}_{\beta}^{\, \lambda(-)}(q)\, {\mathcal{F}}_{ \bar \beta}^{\,\bar \lambda(-)}(\bar q)
\end{eqnarray*}
where
\begin{eqnarray}
\nonumber
\mathcal{F}_{\beta}^{\, \lambda(\pm)}(q)
&=&
\mathcal{F}_{ \beta, { \scriptscriptstyle \rm e}}^{\, \lambda(\pm)}(q)
+
\mathcal{F}_{ \beta, { \scriptscriptstyle \rm o}}^{\, \lambda(\pm)}(q),
\\[12pt]
\nonumber
\mathcal{F}_{ \beta, { \scriptscriptstyle \rm e/o}}^{\, \lambda(\pm)}(q)
&=&
q^{\Delta_\beta-\frac{c}{24}}\,
\sum\limits_{f=0}^\infty\, q^{ f}{F}_{\beta, { \scriptscriptstyle \rm e/o}}^{\, \lambda(\pm), f},
 \\
 \label{blockCoeffR}
 {F}_{ \beta, { \scriptscriptstyle \rm e}}^{\, \lambda(\pm), f}
 &=&
\hspace{-30pt}
\sum\limits_{\begin{array}{c}\scriptstyle {f=|M|+|K|=|N|+|L|}\\[-6pt]
 \scriptstyle {\# K,\#L\in 2\mathbb{N}}
\end{array}}
\hspace{-20pt}
\rho^{(\pm)}_{\scriptscriptstyle \rm RR, e}(w^+_{\beta,MK}, \nu_\lambda,w^+_{\beta,NL})
\Big[ B^f_{c\,\beta} \Big]^{MK,NL},
\\
\nonumber
{F}_{ \beta, { \scriptscriptstyle \rm o}}^{\, \lambda(\pm), f}
&=&
\hspace{-30pt}
\sum\limits_{\begin{array}{c}\scriptstyle {f=|M|+|K|=|N|+|L|}\\[-6pt]
 \scriptstyle {\# K,\# L\in 2\mathbb{N}+1}
\end{array}}
\hspace{-20pt}
\rho^{(\pm)}_{\scriptscriptstyle \rm RR, e}(w^+_{\beta,MK}, \nu_\lambda,w^+_{\beta,NL})
\Big[ B^f_{c\,\beta} \Big]^{MK,NL}.
\end{eqnarray}
The representation for the 1-point toric function of $\tilde \phi_{\lambda,\bar \lambda} $
reads
\begin{eqnarray*}
\langle \tilde \phi_{\lambda,\bar \lambda} \rangle_{\scriptscriptstyle \rm R}
&=&
\sum_{(\beta,\bar\beta)}
iC^{\lambda\,\bar \lambda(+)}_{\beta, \bar \beta}
 \mathcal{F}_{\beta}^{* \lambda(+)}(q)\,  {\mathcal{F}}_{ \bar \beta}^{*\bar  \lambda(+)}(\bar q)
 +
 \sum_{(\beta,\bar\beta)}
iC^{\lambda\,\bar \lambda(-)}_{\beta, \bar \beta}
 \mathcal{F}_{\beta}^{* \lambda(-)}(q)\, {\mathcal{F}}_{ \bar \beta}^{*\bar  \lambda(-)}(\bar q)
\end{eqnarray*}
where
\begin{eqnarray}
\nonumber
\mathcal{F}_{\beta}^{* \lambda(\pm)}(q)
&=&
\mathcal{F}_{ \beta, { \scriptscriptstyle \rm e}}^{* \lambda(\pm)}(q)
+
\mathcal{F}_{ \beta, { \scriptscriptstyle \rm o}}^{* \lambda(\pm)}(q),
\\[12pt]
\nonumber
\mathcal{F}_{ \beta, { \scriptscriptstyle \rm e/o}}^{* \lambda(\pm)}(q)
&=&
q^{\Delta_\beta-\frac{c}{24}}\,
\sum\limits_{f=0}^\infty \, q^{ f}{F}_{\beta, { \scriptscriptstyle \rm e/o}}^{* \lambda(\pm), f},
 \\
 \label{blockCoeffR*}
 {F}_{ \beta, { \scriptscriptstyle \rm e}}^{* \lambda(\pm), f}
 &=&
\hspace{-30pt}
\sum\limits_{\begin{array}{c}\scriptstyle {f=|K|+|M|=|L|+|N|}\\[-6pt]
 \scriptstyle {\# K,\# L\in 2\mathbb{N}}
\end{array}}
\hspace{-20pt}
\rho^{(\pm)}_{\scriptscriptstyle \rm RR, o}(w^+_{\beta,MK}, *\nu_\lambda,w^+_{\beta,NL})
\Big[ B^f_{c\, \beta} \Big]^{MK,NL},
\\
\nonumber
{F}_{ \beta, { \scriptscriptstyle \rm o}}^{* \lambda(\pm), f}
&=&
\hspace{-30pt}
\sum\limits_{\begin{array}{c}\scriptstyle {f=|K|+|M|=|L|+|N|}\\[-6pt]
 \scriptstyle {\# K,\# L\in 2\mathbb{N}+1}
\end{array}}
\hspace{-20pt}
\rho^{(\pm)}_{\scriptscriptstyle \rm RR, o}(w^+_{\beta,MK}, *\nu_\lambda, w^+_{\beta,NL})
\Big[ B^f_{c\, \beta} \Big]^{MK,NL}.
\end{eqnarray}
As in the case of 4-point blocks on the sphere \cite{Suchanek:2010kq} one can show
\begin{eqnarray*}
\mathcal{F}_{ \beta, { \scriptscriptstyle \rm e}}^{\, \lambda(\pm)}(q)
&=&\pm
\mathcal{F}_{ \beta, { \scriptscriptstyle \rm o}}^{\, \lambda(\pm)}(q),
\\
\mathcal{F}_{ \beta, { \scriptscriptstyle \rm e}}^{* \lambda(\pm)}(q)
&=&\mp
\mathcal{F}_{ \beta, { \scriptscriptstyle \rm o}}^{* \lambda(\pm)}(q).
\end{eqnarray*}
Hence
\begin{eqnarray*}
\langle \phi_{\lambda,\bar \lambda} \rangle_{\scriptscriptstyle \rm R}
&=&
\;4\sum_{(\beta,\bar\beta)}
 C^{\lambda\,\bar \lambda(+)}_{\beta, \bar \beta}
 \mathcal{F}_{c\, \beta, { \scriptscriptstyle \rm e}}^{\, \lambda(+)}(q)\,
 {\mathcal{F}}_{c\, \bar \beta, { \scriptscriptstyle \rm e}}^{\,\bar \lambda(+)}(\bar q),
\\
\langle \phi_{\lambda,\bar \lambda} \rangle_{\scriptscriptstyle \rm \widetilde R}
&=&
\;4\sum_{(\beta,\bar\beta)}
 C^{\lambda\,\bar \lambda(-)}_{\beta, \bar \beta}
 \mathcal{F}_{c\, \beta, { \scriptscriptstyle \rm e}}^{\, \lambda(-)}(q)\,
 {\mathcal{F}}_{c\, \bar \beta, { \scriptscriptstyle \rm e}}^{\, \bar \lambda(-)}(\bar q),
  \\
\langle \tilde \phi_{\lambda,\bar \lambda} \rangle_{\scriptscriptstyle \rm R}
&=&
4i\sum_{(\beta,\bar\beta)}
 C^{\lambda\,\bar \lambda(-)}_{\beta, \bar \beta}
 \mathcal{F}_{c\, \beta, { \scriptscriptstyle \rm e}}^{* \lambda(-)}(q)\,
 {\mathcal{F}}_{c\, \bar \beta, { \scriptscriptstyle \rm e}}^{*\bar  \lambda(-)}(\bar q),
\\
\langle \tilde \phi_{\lambda,\bar \lambda} \rangle_{\scriptscriptstyle \rm \widetilde R}
&=&
4i\sum_{(\beta,\bar\beta)}
 C^{\lambda\,\bar \lambda(+)}_{\beta, \bar \beta}
 \mathcal{F}_{c\, \beta, { \scriptscriptstyle \rm e}}^{* \lambda(+)}(q)\,
  {\mathcal{F}}_{c\, \bar \beta, { \scriptscriptstyle \rm e}}^{*\bar \lambda(+)}(\bar q),
\end{eqnarray*}
and it is enough to consider the even blocks alone.

\section{Residues}
\subsection{${\rm \bf NS}$ and $ \widetilde{\rm \bf NS}$   sectors}
The method to derive the recursion relations is essentially the same as in the Virasoro algebra case
\cite{Hadasz:2009db}.
The blocks' coefficients (\ref{blockCoeff})
are polynomials in the external weight $\Delta_\lambda$
and rational functions of the intermediate weight $\Delta$ and the central charge $c.$
The poles in ~$\Delta$ are given by the Kac determinant formula for the NS Verma modules
($r + s\in 2{\mathbb N}$):
\begin{eqnarray}
\label{delta:rs}
\Delta_{rs}
& = &
\frac{1-rs}{4} + \frac{1-r^2}{8}b^2 + \frac{1-s^2}{8}\frac{1}{b^2}\,,\hskip 10mm
c=\frac{3}{2} +3\left(b+\frac{1}{b}\right)^2.
\end{eqnarray}
They are related to the  null states
$$
\ket{\chi_{rs}}=D_{rs} \ket{\Delta_{rs}}.
$$
 in the Verma modules ${\cal V}_{\Delta_{rs}}$.
For a generic value of the central charge the
modules ${\cal V}_{\Delta_{rs}+\frac{rs}{2}}$
 are irreducible and the poles are simple \cite{Hadasz:2006sb}, hence:
\begin{equation}
\label{first:expansion:Delta}
F^{\upodkr\lambda, f}_{ \Delta}
=
{\rm h}^{\upodkr\lambda, f}_{ \Delta}
+
\begin{array}[t]{c}
{\displaystyle\sum} \\[-6pt]
{\scriptscriptstyle
1 < rs \leq 2{f}}
\\[-8pt]
{\scriptscriptstyle
r + s\in 2{\mathbb N}
}
\end{array}
\frac{
R^{\upodkr\lambda, f}_{ rs}
}
{
\Delta-\Delta_{rs}
}\,.
\end{equation}
Following the method of \cite{Hadasz:2006sb} one gets
\begin{eqnarray}
\label{res:NN}
R^{\upodkr\lambda, f}_{ rs}
&=&
 \lim_{\Delta \to \Delta_{rs}} (\Delta - \Delta_{rs}) \,
 F^{\upodkr\lambda, f}_{ \Delta}
 \\ [4pt]
 \nonumber
&=& A_{rs} \, \times
 \hspace*{-40pt}
\begin{array}[t]{c}
{\displaystyle\sum} \\[2pt]
{\scriptstyle
|K|+|M| = |L|+|N| = f-\frac{rs}{2}
}
\end{array}
\hspace*{-30pt}
   \rho^{\podkr}_{\scriptscriptstyle \rm NN}(L_{-M}S_{-K}\chi_{rs} , \podkr\nu_\lambda, L_{-N}S_{-L}\chi_{rs} )
\,
 \left[B^{{f}-\frac{rs}{2}}_{ \Delta_{rs}+\frac{rs}{2}}\right]^{KM,LN}  ,
\end{eqnarray}
where the coefficients $A_{rs}$ are given by:
\begin{equation}
\label{A:rs:1}
\frac{1}{A_{rs}}
=
\lim_{\Delta\to\Delta_{rs}}
\frac{\left\langle\Delta|D_{rs}^\dagger D_{rs}|\Delta\right\rangle}{\Delta - \Delta_{rs}}
.
\end{equation}
The exact formula for $A_{rs}(c)$ was proposed by A.~Belavin and Al.~Zamolodchikov  in \cite{Belavin:2006pv}. It reads
\begin{eqnarray}
\label{A:rs:2}
A_{rs}(c)
=
2^{rs-2}
\prod_{m=1-r}^r
\prod_{n=1-s}^s
\left(m b + {n}{b}^{-1}\right)^{-1}
\end{eqnarray}
where $m+n \in 2{\mathbb Z}, \; (m,n) \neq (0,0),(r,s).$
The corresponding expression in the bosonic case was first conjectured by
Al.~Zamolodchikov \cite{Zamolodchikov:2,Zamolodchikov:3}
and recently proved  by Yanagida \cite{Yanagida:2010qm}.

In order to calculate the residues we shall use  the factorization property
of the 3-point blocks
\cite{Hadasz:2006sb}.
For $ |K|+|L|\in \mathbb{N}$ one has in particular
\begin{eqnarray*}
\rho^{\podkr}_{\scriptscriptstyle \rm NN }(L_{-M}S_{-K}\chi_{rs} , \podkr\nu_{\lambda} , L_{-N}S_{-L}\chi_{rs} ) &=&
\\
&&\hspace{-110pt}
\rho^{\podkr}_{\scriptscriptstyle \rm NN }(L_{-M}S_{-K}\nu_{\Delta_{rs}+rs}, \podkr\nu_{\lambda} , L_{-N}S_{-L}\nu_{\Delta_{rs}+rs}) \
\rho^{\podkr}_{\scriptscriptstyle \rm NN }(\chi_{rs}, \podkr\nu_{\lambda} , \chi_{rs})\ .
\end{eqnarray*}
By the same token
\begin{eqnarray}
\label{fctorization}
\rho^{\podkr}_{\scriptscriptstyle \rm NN }(\chi_{rs} , \podkr\nu_{\lambda} , \chi_{rs} )
&=&
\left\{
\begin{array}{lll}
\displaystyle
\rho^{\podkr}_{\scriptscriptstyle \rm NN }(\chi_{rs}, \podkr\nu_{\lambda} , \nu_{\Delta_{rs}+\frac{rs}{2}}) \
\rho^{\podkr}_{\scriptscriptstyle \rm NN }(\nu_{\Delta_{rs}}, \podkr\nu_{\lambda} , \chi_{rs})
& {\rm for}& \frac{rs}{2} \in {\mathbb N}, \\  [5pt]
\displaystyle
\widetilde
{\displaystyle\rho^{\_\_}_{\scriptscriptstyle \rm NN }}(\chi_{rs}, \podkr\nu_{\lambda} , \nu_{\Delta_{rs}+\frac{rs}{2}}) \
\rho^{\podkr}_{\scriptscriptstyle \rm NN }(\nu_{\Delta_{rs}}, \widetilde{\underline{\hspace*{5pt}}\,\nu_{\lambda}} , \chi_{rs})
& {\rm for}& \frac{rs}{2} \in {\mathbb N}-\frac12    ,
\end{array}
\right.
\end{eqnarray}
where $\widetilde
{\rho^{}_{\scriptscriptstyle \rm NN }}={\rho^{*}_{\scriptscriptstyle \rm NN }},\,
\widetilde
{\rho^{*}_{\scriptscriptstyle \rm NN }}={\rho^{}_{\scriptscriptstyle \rm NN }}$ etc.
The 3-point blocks in the formula above can be expressed in terms of the fusion polynomials
\begin{equation}\label{fusionNS}
 \begin{array}{rcl}
 P^{rs}_c\!\left[^{\Delta_2}_{\Delta_1} \right]&=&
\displaystyle
X^{rs}_{\rm e}(\lambda_1 + \lambda_2)X^{rs}_{\rm e}(\lambda_1 - \lambda_2) ,
\\[10pt]
 P^{rs}_c\!\left[^{\ast\Delta_2}_{\ \Delta_1} \right]&=&
\displaystyle
X^{rs}_{\rm o}(\lambda_1 + \lambda_2)X^{rs}_{\rm o}(\lambda_1 - \lambda_2) ,
\end{array}
\end{equation}
where
\begin{equation}\label{Xpolynomials}
 \begin{array}{rcl}
X^{rs}_{\rm e}(\lambda)&=&
\displaystyle \;
{\prod_{p=1-r}^{r-1} \prod_{q=1-s}^{s-1}}
	\left(\frac{\lambda -  p b- q b^{-1}}{2\sqrt2}\right)  ,
\\[15pt]
X^{rs}_{\rm o}(\lambda)&=&
\displaystyle
{ \prod_{p'=1-r}^{r-1} \prod_{q'=1-s}^{s-1}}
 \left(\frac{\lambda - p' b- q' b^{-1}}{2\sqrt2}\right)    ,
\end{array}
\end{equation}
and the products run over:
\begin{eqnarray}\label{p-q}
p = 1-r + 2k, \quad  q=1-s+2l, &\qquad& k+l \in 2 \mathbb{N} \cup \{0\}, \\
p' = 1-r + 2k, \quad  q'=1-s+2l, &\qquad& k+l \in 2 \mathbb{N} +1 \nonumber.
\end{eqnarray}
Using the relations
\vspace*{-5pt}
\begin{equation}
\nonumber
\begin{array}{rcrllllll}
\rho^{\podkr}_{\scriptscriptstyle \rm NN }(\chi_{rs} , \podkr\nu_{2} , \nu_1 )  &=&
\rho^{\podkr}_{\scriptscriptstyle \rm NN }( \nu_1, \podkr\nu_{2} , \chi_{rs} )  &=&
P^{rs}_{c}\!\left[^{\underline{\hspace*{4pt}}\,\Delta_2}_{\ \Delta_{1}}\right]
& {\rm for}&\ \frac{rs}{2} \in {\mathbb N}, \\ [4pt]
\rho^{\podkr}_{\scriptscriptstyle \rm NN }(\chi_{rs} , \widetilde{\underline{\hspace*{5pt}}\,\nu_{2}} , \nu_1 )  &=&
(-1)^{|\widetilde{\_ \,\nu_2}|}
\rho^{\podkr}_{\scriptscriptstyle \rm NN }(\nu_1 , \widetilde{\underline{\hspace*{5pt}}\,\nu_{2}} , \chi_{rs} )  &=&
P^{rs}_{c}\!\left[^{ \widetilde{\underline{\hspace*{4pt}}\,\Delta_{2}}}_{\ \Delta_{1}}\right]
& {\rm for}&\ \frac{rs}{2} \in {\mathbb N}-\frac12 ,
\end{array}
\end{equation}
one gets
\begin{eqnarray}\label{residuaN}
R^{\upodkr\lambda, f}_{ \Delta} =
A_{rs} \, F^{\upodkr\lambda, f-\frac{rs}{2}}_{ \Delta_{rs}+\frac{rs}{2}}
\times \left\{
\begin{array}{llllll}
P^{rs}_{c}\!\left[^{\ \underline{\hspace*{4pt}}\,\Delta_{\lambda}}_{\Delta_{rs}+\frac{rs}{2}}\right] \,
P^{rs}_{c}\!\left[^{\underline{\hspace*{4pt}}\,\Delta_{\lambda}}_{\ \Delta_{rs}}\right]
&{\rm for}& \frac{rs}{2} \in {\mathbb N}, \\ [4pt]
P^{rs}_{c}\!\left[^{\ \widetilde{\underline{\hspace*{4pt}}\,\Delta_{\lambda}}}_{\Delta_{rs}+\frac{rs}{2}}\right] \,
P^{rs}_{c}\!\left[^{ \underline{\hspace*{4pt}}\,\Delta_{\lambda}}_{\ \Delta_{rs}}\right]\,
\widetilde{\_\,s_{rs}}
&{\rm for}& \frac{rs}{2} \in {\mathbb N}-\frac12  ,
\end{array}
\right.
\end{eqnarray}
where $s_{rs} = 1,\,*s_{rs}=(-1)^{rs}$.

\subsection{${\rm \bf R}$ and $ \widetilde{\rm \bf R}$   sectors}

The blocks' coefficients (\ref{blockCoeffR}),  (\ref{blockCoeffR*})
are polynomials in the external weight $\Delta_\lambda$
and rational functions of $\beta$ and the central charge $c.$
The poles in ~$\beta$ are given by the Kac determinant formula for the
positive parity subspace of an R Verma module. They are located at $\pm\beta_{rs}$
where
\begin{eqnarray}\nonumber
\beta_{rs}\;=\;
\frac{1}{2\sqrt{2}} \left( rb + s\frac{1}{b}\right)
\end{eqnarray}
and $r + s\in 2{\mathbb N}+1$.
The poles are related to the  positive parity null states
$$
\ket{\chi^+_{rs}}=D_{rs} \ket{w^+_{rs}}.
$$
For a generic value of the central charge the
modules ${\cal W}_{\beta_{rs}+\frac{rs}{2}}$
 are irreducible and the poles are simple \cite{Suchanek:2010kq}, hence:
\begin{equation}
 {F}_{ \beta, { \scriptscriptstyle \rm e}}^{\,\upodkr\lambda(\pm), f} =
 h^{ \upodkr\lambda(\pm),f}_{\beta}
 +
\hspace{-10pt}
\sum\limits_{\begin{array}{c}\scriptstyle {1<rs\leqslant 2f}\\[-6pt]
 \scriptstyle {r+s\in 2\mathbb N+1 }
\end{array}}
\hspace{-10pt}\left(
{R^{\,\upodkr\lambda(\pm), f}_{\,rs+} \over \beta - \beta_{rs}} +
{R^{\,\upodkr\lambda(\pm), f}_{\,{rs}-} \over \beta + \beta_{rs}}
\right).
\end{equation}
Calculating the residues in the standard way \cite{Suchanek:2010kq} one gets
\begin{eqnarray*}
R^{\,\upodkr\lambda(\pm), f}_{\,rs\pm}
&=&
\lim_{\beta \to \pm\beta_{rs}} (\beta\mp\beta_{rs}) {F}_{ \beta, { \scriptscriptstyle \rm e}}^{\,\upodkr\lambda(\pm), f}
\\
&=&
\mp\frac{1}{2 \beta_{rs}}
A_{rs}
\\
&\times&
\hspace{-30pt}
\sum\limits_{\begin{array}{c}\scriptstyle {f=|K|+|M|=|L|+|N|}\\[-6pt]
 \scriptstyle {| K|,| L|\in 2\mathbb{N}}
\end{array}}
\hspace{-20pt}
\rho^{(\pm)}_{\scriptscriptstyle \rm RR, f}(L_{-M}S_{-K}D_{rs} w^+_{\beta_{rs}}, \podkr\nu_\lambda, L_{-N}S_{-L} D_{rs} w^+_{\beta_{rs}})
\Big[ B^f_{c\,\beta'_{rs}} \Big]^{MK,NL},
\end{eqnarray*}
where
$$
\beta'_{rs}\;=\;
\frac{(-1)^s}{2\sqrt{2}} \left( rb - s\frac{1}{b}\right)
$$
corresponds to the conformal weight $\Delta_{rs}+{rs\over 2}$
and
the parity of the 3-point block is  $\mathrm{f} = \mathrm{e}$ and $ \mathrm{f} =\mathrm{o}$ in the case of $R^{\, \lambda(\pm), f}_{\,rs\pm}$ and $R^{\,\ast \lambda(\pm), f}_{\,rs\pm}$, respectively.
If we assume the normalization
$$D_{rs} = (L_{-1})^{rs\over 2}+\dots
$$
the coefficient
$
A_{rs}$
is given by formula (\ref{A:rs:2}) with $m+n\in 2 \mathbb{Z} +1$ \cite{Belavin:2006pv}.
For this normalization
the odd null state
$
\chi_{rs}^{-} = {{\rm e}^{i{\pi\over 4}}\over i \beta^{\prime}_{rs}}S_0 \chi_{rs}^+
$
can be expressed as
$
\chi_{rs}^- = D_{rs} w_{rs}^-
$ \cite{Dorrzapf:1999nr}.
Using this observation and the properties of the 3-point blocks one obtains the following
factorization formulae \cite{Suchanek:2010kq}
\begin{eqnarray*}
&&
\hspace{-50pt}
\rho^{(\pm)}_{\scriptscriptstyle \rm RR, e}(L_{-M}S_{-K}D_{rs} w^+_{\beta_{rs}}, \nu_\lambda, L_{-N}S_{-L} D_{rs} w^+_{\beta_{rs}})
\\
&=&
\rho^{(\pm)}_{\scriptscriptstyle \rm RR, e}(L_{-M}S_{-K} w^+_{\beta'_{rs}}, \nu_\lambda, L_{-N}S_{-L} w^+_{\beta'_{rs}})
\rho^{(\pm)}_{\scriptscriptstyle \rm RR, e}(D_{rs} w^+_{\beta_{rs}}, \nu_\lambda, D_{rs} w^+_{\beta_{rs}}),
\\
&&
\hspace{-50pt}
\rho^{(\pm)}_{\scriptscriptstyle \rm RR, o}(L_{-M}S_{-K}D_{rs} w^+_{\beta_{rs}}, \ast \nu_\lambda, L_{-N}S_{-L} D_{rs} w^+_{\beta_{rs}})
\\
&=&
\rho^{(\pm)}_{\scriptscriptstyle \rm RR, o}(L_{-M}S_{-K} w^+_{\beta'_{rs}}, \ast\nu_\lambda, L_{-N}S_{-L} w^+_{\beta'_{rs}})
\rho^{(\pm)}_{\scriptscriptstyle \rm RR, e}(D_{rs} w^+_{\beta_{rs}}, \nu_\lambda, D_{rs} w^+_{\beta_{rs}}),
\\
&&
\hspace{-50pt}
\rho^{(\pm)}_{\scriptscriptstyle \rm RR, e}(D_{rs} w^+_{\beta_{rs}}, \nu_\lambda, D_{rs} w^+_{\beta_{rs}})
\\
&=&
\rho^{(\pm)}_{\scriptscriptstyle \rm RR, e}(D_{rs} w^+_{\beta_{rs}}, \nu_\lambda,w^+_{\beta'_{rs}})
\rho^{(\pm)}_{\scriptscriptstyle \rm RR, e}( w^+_{\beta_{rs}}, \nu_\lambda, D_{rs} w^+_{\beta_{rs}}).
\end{eqnarray*}
In terms of the fusion polynomials
$$
P^{rs}_{c}\!\left[_{\pm \beta}^{\;\Delta_{\lambda}}\right]=
X^{rs}_{\rm e}(2\sqrt{2}\beta \mp \lambda)
X^{rs}_{\rm o}(2\sqrt{2}\beta \pm \lambda)
$$
one then has
\begin{eqnarray*}
\rho^{(\pm)}_{\scriptscriptstyle \rm RR, e}( w^+_{\beta}, \nu_\lambda, D_{rs} w^+_{\beta_{rs}})
&=&
\rho^{(\pm)}_{\scriptscriptstyle \rm RR, e}(D_{rs} w^+_{\beta_{rs}}, \nu_\lambda,  w^+_{\beta})
\;=\;
P^{rs}_{c}\!\left[_{\pm \beta}^{\;\Delta_{\lambda}}\right]
\end{eqnarray*}
and
\begin{eqnarray*}
R^{\,\upodkr\lambda(\pm), f}_{\,rs\pm}
&=&
\mp\frac{1}{2 \beta_{rs}}
A_{rs}
P^{rs}_{c}\!\left[_{\pm \beta'_{rs}}^{\;\Delta_{\lambda}}\right]
P^{rs}_{c}\!\left[_{\pm \beta_{rs}}^{\;\Delta_{\lambda}}\right]
{F}_{ \beta'_{rs}, { \scriptscriptstyle \rm e}}^{\,\upodkr\lambda(\pm), f-{rs\over 2}}.
\end{eqnarray*}
Since the residues at $\pm \beta_{rs}$ differ by sign
one simply gets
\begin{equation}
\label{ramec}
 {F}_{ \beta, { \scriptscriptstyle \rm e}}^{\,\upodkr\lambda(\pm), f} =
 h^{\upodkr\lambda(\pm),f}_{\beta}
 +
\hspace{-10pt}
\sum\limits_{\begin{array}{c}\scriptstyle {1<rs\leqslant 2f}\\[-6pt]
 \scriptstyle {r+s\in 2\mathbb N+1 }
\end{array}}
\hspace{-10pt}
{A_{rs}
P^{rs}_{c}\!\left[_{\pm \beta'_{rs}}^{\;\Delta_{\lambda}}\right]
P^{rs}_{c}\!\left[_{\pm \beta_{rs}}^{\;\Delta_{\lambda}}\right] \over \Delta_\beta - \Delta_{rs}}
{F}_{ \beta'_{rs}, { \scriptscriptstyle \rm e}}^{\,\upodkr\lambda(\pm), f-{rs\over 2}}.
\end{equation}
Let us note that  recursions for the blocks ${F}_{ \beta, { \scriptscriptstyle \rm e}}^{\,\lambda(\pm), f}$ and  ${F}_{ \beta, { \scriptscriptstyle \rm e}}^{\,\ast\lambda(\pm), f}$ are very similar.
 The only difference is  the function $ h^{\upodkr\lambda(\pm),f}_{\beta} $ which we determine in the next section.

\section{Large $\Delta$ asymptotics}
In order to complete the derivation of recurrence relations one needs
the large $\Delta$ asymptotics of  conformal blocks.
Their rigorous calculation turned out however to be more difficult that in the Virasoro algebra case \cite{Hadasz:2009db}.
The method presented in this section  is based on  properties of the Gram matrix
and the matrix elements of the chiral vertex operators, collected in Propositions 1 -- 4.
Their proofs are given in Appendices A and B.

\subsection{${\rm \bf NS}$ and $ \widetilde{\rm \bf NS}$   sectors}
Let ${\cal B}_f$ denotes the standard basis of level $f$ subspace of the NS Verma module:
\begin{eqnarray}
\label{fullbasis}
{\cal B}_f&=&
\{L_{-M}S_{-K} \nu_\Delta \,:\, |M|+|K|=f \}.
\end{eqnarray}
It is convenient to use a simplified notation for elements of this basis
\[
{\cal B}_f=\{u_i\}_{i=1}^{\dim {\cal B}_f}.
\]

\noindent{\bf Proposition 1}

\noindent {\it Let $Q$ be a polynomial in $\Delta$ and let ${\rm deg}_{\Delta}\,Q$ denotes its degree. Then:
\begin{enumerate}
\item
for any $u_i= L_{-M}S_{-K}\nu_\Delta \in {\cal B}_f$:
$$
{\rm deg}_{\Delta} \langle u_i | u_i \rangle = \# M +\#K;
$$
\item
for any $u_i,u_j \in {\cal B}_f, u_i\neq u_j$:
$$
{\rm deg}_{\Delta} \langle u_i | u_j \rangle <
{\rm max}
\left\{
{\rm deg}_{\Delta} \langle u_i | u_i\rangle
,
{\rm deg}_{\Delta} \langle u_j | u_j\rangle
\right\};
$$
\item
the product of the diagonal terms is the only highest degree term in the determinant of the Gram matrix
with respect to the base ${\cal B}_f$ i.e.
$$
{\rm deg}_{\Delta} \left(\det \Big[\langle u_i, u_j\rangle\Big] - \prod\limits_{i=1}^{\dim  {\cal B}_f} \langle u_i, u_i\rangle\right)
<{\rm deg}_{\Delta} \det \Big[\langle u_i, u_j\rangle\Big].
$$
\end{enumerate}
}

\noindent {\bf Proposition 2}

{\it
\noindent
For any $u_i,u_j \in {\cal B}_f,\, u_i\neq u_j$:
\begin{eqnarray*}
{\rm deg}_{\Delta} \,
\rho_{\scriptscriptstyle \rm NN }(u_i, \nu_{\lambda} , u_j)
&<&
{\rm max}
\left\{
{\rm deg}_{\Delta} \langle u_i | u_i\rangle ,
{\rm deg}_{\Delta} \langle u_j | u_j\rangle
\right\},
\\
{\rm deg}_{\Delta} \,
\rho^*_{\scriptscriptstyle \rm NN }(u_i, *\nu_{\lambda} , u_j)
&<&
{\rm max}
\left\{
{\rm deg}_{\Delta} \langle u_i | u_i\rangle ,
{\rm deg}_{\Delta} \langle u_j | u_j\rangle
\right\}.
\end{eqnarray*}

}

\noindent By Proposition 2, for off diagonal elements 
\begin{equation}
\label{eqqq}
{\rm deg}_{\Delta}\,\rho^{\upodkr}_\NN(u_i,\podkr\nu_\lambda,u_j) < {\rm deg}_{\Delta} \langle u_i | u_i\rangle
\;\;\;{\rm or}\;\;\;
{\rm deg}_{\Delta}\,\rho^{\upodkr}_\NN(u_i,\podkr\nu_\lambda,u_j) < {\rm deg}_{\Delta} \langle u_j | u_j\rangle.
\end{equation}
Suppose the first inequality holds.
The minor $M_{ji}$  of the Gram matrix can be represented as
\begin{eqnarray*}
M_{ji} & = &
\sum\limits_{\tau} {\rm sgn}(\tau)
\langle u_1|u_{\tau(1)}\rangle \ldots {\langle u_{i-1}|u_{\tau(i-1)}\rangle}\langle u_{i+1}|u_{\tau(i+1)}\rangle\ldots 
\end{eqnarray*}
where the sum runs over permutations $\tau$ such that $\tau(i) = j.$ By Proposition 1:
\[
{\rm deg}_\Delta\langle u_k|u_{\tau(k)}\rangle \; \leqslant \; {\rm deg}_\Delta\,\langle u_k|u_k\rangle
\]
hence, for every permutation $\tau:$
\begin{eqnarray*}
{\rm deg}_\Delta \big(\langle u_1|u_{\tau(1)}\rangle \ldots {\langle u_{i-1}|u_{\tau(i-1)}\rangle}\langle u_{i+1}|u_{\tau(i+1)}\rangle\ldots\big)
& \leqslant &
\sum\limits_{k\neq j}{\rm deg}_\Delta\langle u_k|u_k\rangle.
\end{eqnarray*}
Taking into account the first inequality of (\ref{eqqq}) one thus gets
\[
{\rm deg}_\Delta\big(\rho^{\upodkr}_\NN(u_i,\podkr\nu_\lambda,u_j)M_{ji}\big) < \sum\limits_{k}{\rm deg}_\Delta\langle u_k|u_k\rangle = {\rm deg}_\Delta \det \Big[\langle u_i, u_j\rangle\Big].
\]
 and
\begin{equation}
\lim\limits_{\Delta\to\infty}\rho^{\upodkr}_\NN(u_i,\podkr\nu_\lambda,u_j)B^{ij}_f
=
(-1)^{i+j}\lim\limits_{\Delta\to\infty}\frac{\rho^{\upodkr}_\NN(u_i,\podkr\nu_\lambda,u_j) M_{ji}}{\det \Big[\langle u_i, u_j\rangle\Big]}
= 0.
\end{equation}
If the second inequality of  (\ref{eqqq}) holds one follows the same reasoning with a different minor representation:
\begin{eqnarray*}
M_{ji} & = &
\sum\limits_{\tau} {\rm sgn}(\tau)
\langle u_{\tau(1)}| u_1\rangle \ldots {\langle u_{\tau(j-1)}| u_{j-1}\rangle}\langle u_{\tau(j+1)}| u_{j+1}\rangle\ldots 
\end{eqnarray*}
where the sum runs over permutations of $\tau$ such that $\tau(j) = i.$
Thus for  $i\neq j:$
\begin{equation}
\label{as1}
\lim\limits_{\Delta\to\infty}\rho^{\upodkr}_\NN(u_i,\podkr\nu_\lambda,u_j)B^{ij}
= 0.
\end{equation}
One easily shows (see the proof of Proposition 2 in Appendix A)
that the term of the highest $\Delta$ degree in $\rho^{\upodkr}_\NN(u_i,\podkr\nu_\lambda,u_i)$
is equal to $\langle u_i|u_i\rangle.$
Hence
\begin{equation}
\label{as2}
\lim\limits_{\Delta\to\infty}\rho^{\upodkr}_\NN(u_i,\podkr\nu_\lambda,u_i)B^{ii}
= 1.
\end{equation}
(There is no summation over repeated indices in formulae (\ref{as1}), (\ref{as2}).)
One  finally gets
\begin{eqnarray}
\nonumber
{\rm h}^{\upodkr\lambda, f}_{ \Delta}
&=&
\lim\limits_{\Delta\to\infty}F^{\upodkr\lambda, f}_{ \Delta}
=
\lim\limits_{\Delta\to\infty}\Big(\sum\limits_{i,j = 1}^{{\rm dim}\,{\cal B}_f}\rho^{\podkr}_\NN(u_i,\podkr\nu_\lambda,u_j) B^{ij}_f \Big)
\\
\label{as3}
&=&
\sum\limits_{i=1}^{{\rm dim}\,{\cal B}_f} 1
=
{\rm dim}\,{\cal B}_f=p_{\scriptscriptstyle \rm NS }(f)
\end{eqnarray}
where $p_{\scriptscriptstyle \rm NS }(f)$  is defined by the generating function
\begin{eqnarray*}
\sum\limits_{f\in \frac12{\mathbb N}\cup\{0\}}\hskip -7pt p_{\scriptscriptstyle \rm NS }(f) q^f
& = &
\prod\limits_{n=1}^\infty\ {\;\;1+q^{n-\frac12}\!\!\!\!\over 1-q^n}
\;=\;
q^{\frac{c}{24}-\Delta}\chi^\Delta_{\scriptscriptstyle \rm NS }(q)
\end{eqnarray*}
and $\chi^\Delta_{\scriptscriptstyle \rm NS }(q)$ is the character of the NS Verma module
\cite{Goddard:1986ee,Matsuo:1986vc,Kastor:1986ig,Cappelli:1986ed}
$$
\chi^\Delta_{\scriptscriptstyle \rm NS }(q)
=
{\rm Tr}_{\scriptscriptstyle \rm NS } \,q^{L_0 - \frac{c}{24}}
=
q^{{\Delta-\frac{c}{24}+{1\over 16}}} \, \eta(q)^{-\frac32} \, \sqrt{\theta_3(q^{1/2})}\,.
$$
For ``twisted'' blocks 
(\ref{tildeblocks}) asymptotic (\ref{as3}) implies
$$
\widetilde{{\rm h}}^{\upodkr\lambda, f}_{ \Delta}
=
(-1)^{2f}\lim\limits_{\Delta\to\infty}{F}^{ \upodkr\lambda, f}_{ \Delta}
=
(-1)^{2f}p_{{{\scriptscriptstyle \rm NS }}}(f)
=p_{\widetilde{{\scriptscriptstyle \rm NS }}}(f).
$$
The generating function for $p_{\widetilde{ \scriptscriptstyle \rm NS }}(f)$ takes the form
\begin{eqnarray*}
\sum\limits_{f\in \frac12{\mathbb N}\cup\{0\}}\hskip -7pt p_{ \widetilde{\scriptscriptstyle \rm NS} }(f) q^f
& = &
\prod\limits_{n=1}^\infty\ {\;\;1-q^{n-\frac12}\!\!\!\!\over 1-q^n}
\;=\;
q^{\frac{c}{24}-\Delta}\chi^\Delta_{\widetilde{ \scriptscriptstyle \rm NS} }(q)
\end{eqnarray*}
where  $\chi^\Delta_{\widetilde{ \scriptscriptstyle \rm NS }}(q)$ is a modified character
\cite{Matsuo:1986vc,Kastor:1986ig,Cappelli:1986ed}
$$
\chi^\Delta_{ \widetilde{\scriptscriptstyle \rm NS }}(q)
=
{\rm Tr}_{\scriptscriptstyle \rm NS } \,(-)^F q^{L_0 - \frac{c}{24}}
=
q^{{\Delta-\frac{c}{24}+{1\over 16}}} \, \eta(q)^{-\frac32} \, \sqrt{\theta_4(q^{1/2})}.
$$

\subsection{${\rm \bf R}$ and $ \widetilde{\rm \bf R}$   sectors}
In order to calculate the large $\Delta_\beta$ behavior of the Ramond toric blocks we shall chose a special basis
${\cal B}_f$ of level $f$ even subspace of the Ramond Verma module. It is defined by:
\begin{eqnarray}
\nonumber
\label{fullbasis:R}
{\cal B}_f&=&{\cal B}^+_f\cup {\cal B}^-_f,
\\
{\cal B}_f^+&=&
\{L_{-M}S_{-K} w_\beta^+ \,:\, |M|+|K|=f ,\ \#K \in 2\mathbb{N}\cup\{0\}\},
\\
\nonumber
{\cal B}_f^-&=&
\{L_{-M}S_{-K} w_\beta^- \,:\, |M|+|K|=f ,\ \#K \in 2\mathbb{N} + 1\},
\end{eqnarray}
where the string of generators $S_{-K}$ does not include $S_0.$
We shall also use a simplified notation for elements of the basis above:
$$
{\cal B}_f=\{u_i\}_{i=1}^{\dim {\cal B}_f},
\hskip 1cm
{\cal B}^\pm_f=\{u^\pm_k\}_{k=1}^{\dim {\cal B}^\pm_f}.
$$
As it is shown in Appendix B the subsets ${\cal B}^\pm_f$ are composed of the same number of elements,
\begin{equation}
\label{equal:dimensionality:of:pm:subspaces}
{\rm dim}\,{\cal B}_f^+= {\rm dim}\,{\cal B}_f^-.
\end{equation}

\noindent
{\bf Proposition 3}

\noindent
{\it  Let $Q$ be a polynomial in $\beta$ and let ${\rm deg}_{\beta}\,Q$ denotes its degree. Then{\em :}
\begin{enumerate}
\item
for any $u_k^+ = L_{-M}S_{-K}w_\beta^+ \in {\cal B}^+_f$ and for any $u_l^- = L_{-M}S_{-K}w_\beta^- \in {\cal B}^-_f:$
$$
{\rm deg}_\beta\langle u_k^+ | u_k^+ \rangle  = 2(\#M + \# K),
\hskip 1cm
{\rm deg}_\beta\langle u_l^- | u_l^- \rangle  = 2(\#M + \# K);
$$
\item
for any
$u^+_k\in {\cal B}^+_n$, $u^-_l\in {\cal B}^-_n:$
$$
{\rm deg}_\beta\langle u^+_k | u^-_l \rangle <
{\rm min}
\left\{
{\rm deg}_\beta\langle u^-_l| u^-_l \rangle,
{\rm deg}_\beta\langle u^+_k | u^+_k \rangle
\right\};
$$
\item
for any
$u^\pm_k,u^\pm_l\in {\cal B}^\pm_n$, $u^\pm_k\neq u^\pm_l:$
$$
{\rm deg}_\beta\langle u^\pm_k | u^\pm_l \rangle <
{\rm max}
\left\{
{\rm deg}_\beta\langle u^\pm_l | u^\pm_l \rangle,
{\rm deg}_\beta\langle u^\pm_k | u^\pm_k \rangle
\right\};
$$
\item
the product of the diagonal terms is the only highest degree term in the determinant of the Gram matrix
with respect to the base ${\cal B}_n$ i.e.
$$
{\rm deg}_\beta\left(\det \Big[\langle u_i, u_j\rangle\Big] - \prod\limits_{i=1}^{\dim  {\cal B}_n} \langle u_i, u_i\rangle\right)
<{\rm deg}_\beta\det \Big[\langle u_i, u_j\rangle\Big].
$$
\end{enumerate}
}

Let us recall that matrix elements of  arbitrary chiral vertex operators
$
V(\nu_{\lambda}), V(*\nu_{\lambda})
$
 between even states $u_i, u_j \in {\cal W}_\beta$ can be decomposed as \cite{Chorazkiewicz:2011zd}:
\begin{eqnarray}
\label{dec}
\langle u_i |V(\nu_{\lambda}) | u_j \rangle &=&
\\
\nonumber
&&\hspace{-50pt}
 \rho^{++}_\RRe(u_i,\nu_\lambda,u_j) \langle w^+_\beta |V(\nu_{\lambda}) | w^+_\beta \rangle
+\rho^{--}_\RRe(u_i,\nu_\lambda,u_j) \langle w^-_\beta |V(\nu_{\lambda}) | w^-_\beta \rangle,
\\
\label{dec*}
\langle u_i |V(*\nu_{\lambda}) | u_j \rangle &=&
\\
\nonumber
&&\hspace{-50pt}
 \rho^{+-}_\RRo(u_i,*\nu_\lambda,u_j) \langle w^+_\beta |V(\nu_{\lambda}) | w^-_\beta \rangle
+\rho^{-+}_\RRo(u_i,*\nu_\lambda,u_j) \langle w^-_\beta |V(\nu_{\lambda}) | w^+_\beta \rangle.
\end{eqnarray}
The decompositions above can be seen as defining the forms $\rho^{\pm\pm}_\RRe,\ \rho^{\pm\mp}_\RRo$.
They are related to
3-point blocks (\ref{chiral:vertex:RNR})  by
\begin{eqnarray}
\nonumber
\rho^{(\pm)}_{\scriptscriptstyle \rm RR, e}(u_i^{\pm}, \nu_\lambda,u_j^{\pm})
& = &
\rho^{++}_{\scriptscriptstyle \rm RR, e}(u_i^{\pm}, \nu_\lambda,u_j^{\pm})
\pm
\rho^{--}_{\scriptscriptstyle \rm RR, e}(u_i^{\pm}, \nu_\lambda,u_j^{\pm}),
\\[-7pt]
\label{old:forms:rho}
\\[-7pt]
\nonumber
\rho^{(\pm)}_{\scriptscriptstyle \rm RR, o}(u_i^{\pm}, *\nu_\lambda,u_j^{\pm})
& = &
\rho^{+-}_{\scriptscriptstyle \rm RR, o}(u_i^{\pm}, *\nu_\lambda,u_j^{\pm})
\pm
i\rho^{-+}_{\scriptscriptstyle \rm RR, o}(u_i^{\pm}, *\nu_\lambda,u_j^{\pm}).
\end{eqnarray}
There holds:

\noindent
{\bf Proposition 4}

\noindent Let
$
{\rm deg}_\beta \langle w^\pm_\beta |V(\nu_{\lambda}) | w^\pm_\beta \rangle = {\rm deg}_\beta \langle w^\pm_\beta |V(\nu_{\lambda}) | w^\mp_\beta \rangle = 0.
$
Then:
{\em \begin{enumerate}
\item
for any $u^+_k,u^+_l \in {\cal B}_f,\, u^+_k\neq u^+_l:$
\begin{eqnarray*}
{\rm deg}_\beta\,
\rho^{++}_\RRe(u^+_k,\nu_\lambda,u^+_l)
&<&
{\rm max}
\left\{
{\rm deg}_\beta\langle u^+_k | u^+_k\rangle ,
{\rm deg}_\beta\langle u^+_l | u^+_l\rangle
\right\},
\\
{\rm deg}_\beta\,\beta^{-1}
\rho^{+-}_\RRo(u^+_k,*\nu_\lambda,u^+_l)
&<&
{\rm max}
\left\{
{\rm deg}_\beta\langle u^+_k | u^+_k\rangle,
{\rm deg}_\beta\langle u^+_l | u^+_l\rangle
\right\},
\\
{\rm deg}_\beta\,\beta^{-1}
\rho^{-+}_\RRo(u^+_k,*\nu_\lambda,u^+_l)
&<&
{\rm max}
\left\{
{\rm deg}_\beta\langle u^+_k | u^+_k\rangle,
{\rm deg}_\beta\langle u^+_l | u^+_l\rangle
\right\};
\end{eqnarray*}
\item
for any $u^-_k,u^-_l \in {\cal B}_f,\, u^-_k\neq u^-_l:$
\begin{eqnarray*}
{\rm deg}_\beta\,
\rho^{--}_\RRe(u^-_k,\nu_\lambda,u^-_l)
&<&
{\rm max}
\left\{
{\rm deg}_\beta\langle u^-_k | u^-_k\rangle ,
{\rm deg}_\beta\langle u^-_l | u^-_l\rangle
\right\},
\\
{\rm deg}_\beta\,\beta^{-1}
\rho^{+-}_\RRo(u^-_k,*\nu_\lambda,u^-_l)
&<&
{\rm max}
\left\{
{\rm deg}_\beta\langle u^-_k | u^-_k\rangle,
{\rm deg}_\beta\langle u^-_l | u^-_l\rangle
\right\},
\\
{\rm deg}_\beta\,\beta^{-1}
\rho^{-+}_\RRo(u^-_k,*\nu_\lambda,u^-_l)
&<&
{\rm max}
\left\{
{\rm deg}_\beta\langle u^-_k | u^-_k\rangle,
{\rm deg}_\beta\langle u^-_l | u^-_l\rangle
\right\};
\end{eqnarray*}
\item
for any $u^\pm_k,u^\pm_l \in {\cal B}_f:$
\begin{eqnarray*}
{\rm deg}_\beta\,
\rho^{++}_\RRe(u^-_k,\nu_\lambda,u^-_l)
&<&
{\rm min}
\left\{
{\rm deg}_\beta\langle u^-_k | u^-_k\rangle ,
{\rm deg}_\beta\langle u^-_l | u^-_l\rangle
\right\},
\\
{\rm deg}_\beta\,
\rho^{--}_\RRe(u^+_k,\nu_\lambda,u^+_l)
&<&
{\rm min}
\left\{
{\rm deg}_\beta\langle u^+_k | u^+_k\rangle ,
{\rm deg}_\beta\langle u^+_l | u^+_l\rangle
\right\};
\end{eqnarray*}
\item
for any $u^\pm_k,u^\pm_l \in {\cal B}_f:$
\begin{eqnarray*}
{\rm deg}_\beta\,
\rho^{++}_\RRe(u^\pm_k,\nu_\lambda,u^\mp_l)
&<&
{\rm min}
\left\{
{\rm deg}_\beta\langle u^\pm_k | u^\pm_k\rangle ,
{\rm deg}_\beta\langle u^\mp_l | u^\mp_l\rangle
\right\},
\\
{\rm deg}_\beta\,
\rho^{--}_\RRe(u^\pm_k,\nu_\lambda,u^\mp_l)
&<&
{\rm min}
\left\{
{\rm deg}_\beta\langle u^\pm_k | u^\pm_k\rangle ,
{\rm deg}_\beta\langle u^\mp_l | u^\mp_l\rangle
\right\},
\\
{\rm deg}_\beta\,\beta^{-1}
\rho^{+-}_\RRo(u^\pm_k,*\nu_\lambda,u^\mp_l)
&<&
{\rm min}
\left\{
{\rm deg}_\beta\langle u^\pm_k | u^\pm_k\rangle,
{\rm deg}_\beta\langle u^\mp_l | u^\mp_l\rangle
\right\},
\\
{\rm deg}_\beta\,\beta^{-1}
\rho^{-+}_\RRo(u^\pm_k,*\nu_\lambda,u^\mp_l)
&<&
{\rm min}
\left\{
{\rm deg}_\beta\langle u^\pm_k | u^\pm_k\rangle,
{\rm deg}_\beta\langle u^\mp_l | u^\mp_l\rangle
\right\}.
\end{eqnarray*}
\end{enumerate}
}

\noindent It follows from Proposition 4 that for any $u_i, u_j \in {\cal B}_f,\ u_i \neq u_j:$
\[
{\rm deg}_\beta\,\rho^{\pm\pm}_\RRe(u_i,\nu_\lambda,u_j) < {\rm deg}_\beta\,\langle u_i|u_i\rangle
\hskip 5mm {\rm or}\hskip 5mm
{\rm deg}_\beta\,\rho^{\pm\pm}_\RRe(u_i,\nu_\lambda,u_j) < {\rm deg}_\beta\,\langle u_j|u_j\rangle
\]
and
\[
{\rm deg}_\beta\,\beta^{-1}\rho^{\pm\mp}_\RRo(u_i,*\nu_\lambda,u_j) < {\rm deg}_\beta\,\langle u_i|u_i\rangle
\hskip 5mm {\rm or}\hskip 5mm
{\rm deg}_\beta\,\beta^{-1}\rho^{\pm\mp}_\RRo(u_i,*\nu_\lambda,u_j) < {\rm deg}_\beta\,\langle u_j|u_j\rangle.
\]
Following the same steps as in the previous subsection we thus get for $i\neq j:$
\begin{eqnarray}
\nonumber
\lim\limits_{\beta\to\infty}\rho^{\pm\pm}_\RRe(u_i,\nu_\lambda,u_j)B^{ij}  & = & 0,
\\[-8pt]
\label{Ramond:non:diagonal}
\\[-8pt]
\nonumber
\lim\limits_{\beta\to\infty}\beta^{-1}\rho^{\pm\mp}_\RRo(u_i,\nu_\lambda,u_j) B^{ij}  & = & 0.
\end{eqnarray}
Let
$$
\left[
\begin{array}{cc}
B^{i j}_{++} & B^{ij}_{+-}  \\
B^{ij}_{-+} & B^{ij}_{--}
\end{array}
\right]
$$
be the matrix inverse to the Gram matrix
$$
\left[
\begin{array}{cc}
\langle u^+_i|u^+_j \rangle & \langle u^+_i|u^-_j \rangle  \\
\langle u^-_i|u^+_j \rangle & \langle u^-_i|u^-_j \rangle
\end{array}
\right].
$$
By Proposition 4, for any $u_k^{\pm} \in {\cal B}_f^\pm:$
\[
{\rm deg}_\beta\,\rho^{\mp\mp}_\RRe(u_k^{\pm},\nu_\lambda,u_k^{\pm}) < {\rm deg}\,\langle u_k^{\pm},u_k^{\pm}\rangle
\]
and we also get
\begin{eqnarray}
\label{Ramond:vanishing:diagonal}
\lim\limits_{\beta\to\infty}\rho^{\mp\mp}_\RRe(u_k^{\pm},\nu_\lambda,u_k^{\pm}))B^{kk}_{\pm\pm}  & = & 0.
\end{eqnarray}
Since
\begin{eqnarray}
\nonumber
\rho^{+-}_\RRo(w^+_\beta,*\nu_\lambda,w^+_\beta) & = & \rho^{+-}_\RRo(w_\beta^-,*\nu_\lambda,w_\beta^-) \, = \,
i\beta {\rm e}^{-i{\pi\over 4}}\;=\;\beta{\rm e}^{+i{\pi\over 4}},
\\[-7pt]
\label{starrho:beta:dependence}
\\[-7pt]
\nonumber
\rho^{-+}_\RRo(w_\beta^+,*\nu_\lambda,w_\beta^+) & = &\rho^{-+}_\RRo(w_\beta^-,*\nu_\lambda,w_\beta^-)  \,=\,  i\beta {\rm e}^{+i{\pi\over 4}}\;=\; - \beta{\rm e}^{-i{\pi\over 4}},
\end{eqnarray}
the leading terms in the large $\beta$ limit take the form:
\begin{eqnarray*}
\rho^{\pm\pm}_\RRe(u_k^{\pm},\nu_\lambda,u_k^{\pm}) & = & \langle u_k^\pm|u_k^\pm\rangle + \ldots,
\\[2pt]
\rho^{+-}_\RRo(u_k^\pm,*\nu_\lambda,u_k^\pm) & =& \beta{\rm e}^{i{\pi\over 4}} \langle u_k^\pm|u_k^\pm\rangle + \ldots,
\\[2pt]
\rho^{-+}_\RRo(u_k^\pm,*\nu_\lambda,u_k^\pm) & =& -\beta{\rm e}^{-i{\pi\over 4}} \langle u_k^\pm|u_k^\pm\rangle + \ldots.
\end{eqnarray*}
This yields
\begin{eqnarray}
\nonumber
\lim\limits_{\beta\to\infty} \rho^{\pm\pm}_\RRe(u_k^{\pm},\nu_\lambda,u_k^{\pm}) B^{kk}_{\pm\pm}  & = & 1,
\\[2pt]
\label{Ramond:nonvanishing:diagonal}
\lim\limits_{\beta\to\infty} \beta^{-1}\rho^{+-}_\RRo(u_k^\pm,*\nu_\lambda,u_k^\pm) & =& {\rm e}^{i{\pi\over 4}},
\\[2pt]
\nonumber
\lim\limits_{\beta\to\infty} \beta^{-1}\rho^{-+}_\RRo(u_k^\pm,*\nu_\lambda,u_k^\pm) & =& -{\rm e}^{-i{\pi\over 4}}.
\end{eqnarray}
For forms (\ref{old:forms:rho}), equations (\ref{Ramond:vanishing:diagonal}) and (\ref{Ramond:nonvanishing:diagonal}) give
\begin{equation}
\label{Ramond:non:zero:limits}
\begin{array}{llllllll}
\lim\limits_{\beta\to\infty}\rho_\RRe^{(\pm)}(u_k^+,\nu_\lambda,u_k^+)B^{kk}_{++} = 1,
\\[10pt]
\lim\limits_{\beta\to\infty}\rho_\RRe^{(\pm)}(u_k^-,\nu_\lambda,u_k^-)B^{kk}_{--} = \pm 1,
\\[10pt]
\lim\limits_{\beta\to\infty}\beta^{-1}\rho_\RRo^{(+)}(u_k^\pm,*\nu_\lambda,u_k^\pm)B^{kk}_{\pm\pm} =0 ,
\\[10pt]
\lim\limits_{\beta\to\infty}\beta^{-1}\rho_\RRo^{(-)}(u_k^\pm,*\nu_\lambda,u_k^\pm)B^{kk}_{\pm\pm} = 2{\rm e}^{\frac{i\pi}{4}}.
\end{array}
\end{equation}
Using (\ref{Ramond:non:diagonal}) and (\ref{Ramond:non:zero:limits}) we finally get (the case $f = 0$ is special as ${\cal B}^-_0 = \varnothing$):
\begin{eqnarray}
\label{Rasymptot:1}
h^{\lambda(+),f}_{\beta}
& = &
\lim_{\beta\to\infty}{F}_{ \beta, { \scriptscriptstyle \rm e}}^{\lambda(+),f}
=
\lim\limits_{\beta\to\infty}\sum\limits_{i=1}^{{\rm dim}\,{\cal B}_f}\rho_\RRe^{(+)}(u_i,\nu_\lambda,u_i)B^{ii}
\\
\nonumber
& = &
\lim\limits_{\beta\to\infty}\sum\limits_{k=1}^{{\rm dim}\,{\cal B}_f^{\pm}}\left(\rho_\RRe^{(+)}(u_k^+,\nu_\lambda,u_k^+)B^{kk}_{++} + \rho_\RRe^{(+)}(u_k^-,\nu_\lambda,u_k^-)B^{kk}_{--}\right)
\\
\nonumber
&=&
{\rm dim}\,{\cal B}_f
=
p_{\scriptscriptstyle\rm R}(f),
\end{eqnarray}
where $p_{\scriptscriptstyle\rm R}(f)$ can be computed by means of generating function
\begin{eqnarray*}
\sum\limits_{q=0}^{\infty}p_{\scriptscriptstyle\rm R}(f)q^f
& = &
\prod\limits_{n=1}^\infty\frac{1+q^n}{1-q^n}
=
{\textstyle {1\over 2}}q^{\frac{c}{24}-\Delta}\chi^\Delta_{\scriptscriptstyle \rm R }(q)
\end{eqnarray*}
and $\chi^\Delta_{\scriptscriptstyle \rm R}(q)$ is the character of the Ramond Verma module
\cite{Goddard:1986ee,Matsuo:1986vc,Kastor:1986ig,Cappelli:1986ed}
$$
\chi^\Delta_{\scriptscriptstyle \rm R }(q)
=
{\rm Tr}_{\scriptscriptstyle \rm R } \,q^{L_0 - \frac{c}{24}}
=
q^{{\Delta-\frac{c}{24}}} \, \eta(q)^{-\frac32} \, \sqrt{2\,\theta_2(q^{1/2})}\,.
$$
Similarly:
\begin{eqnarray}
\nonumber
h^{\lambda(-),f}_{\beta}
& = &
\lim_{\beta\to\infty}{F}_{ \beta, { \scriptscriptstyle \rm e}}^{\lambda(-),f}
=
\delta_{f,0},
\\
\label{Rasymptot:2}
\beta^{-1} h^{*\lambda(+),f}_{\beta}
& = &
\lim_{\beta\to\infty}\beta^{-1}{F}_{ \beta, { \scriptscriptstyle \rm e}}^{*\lambda(+),f}
=
0,
\\
\nonumber
\beta^{-1} h^{*\lambda(-),f}_{\beta}
& = &
\lim_{\beta\to\infty}\beta^{-1}{F}_{ \beta, { \scriptscriptstyle \rm e}}^{\lambda(-),f}
=
2\,{\rm e}^{\frac{i\pi}{4}}p_{\scriptscriptstyle\rm R}(f).
\end{eqnarray}

\section{Elliptic blocks and recurrence relations}

\subsection{${\rm \bf NS}$ and $ \widetilde{\rm \bf NS}$   sectors}

The large $\Delta$ asymptotic (\ref{as3}) suggests the following definition of the elliptic blocks in ${\rm \bf NS}$ sector:
\begin{equation}
\label{elliptic blocks NS}
\begin{array}{rcl}
\mathcal{F}_{ \Delta}^{\upodkr\lambda}(q)&=&
q^{{\Delta-\frac{c}{24}+{1\over 16}}} \, \eta(q)^{-\frac32} \, \sqrt{\theta_3(q^{1/2})} \,
\mathcal{H}_{ \Delta}^{\upodkr\lambda}(q)\ ,
\\[5pt]
\mathcal{H}_{ \Delta}^{\upodkr\lambda}(q)
    &=&
\sum_{f\in {1\over 2}\mathbb{N}\cup\{0\}} q^{f} \,
    H^{\upodkr\lambda, f}_{ \Delta}
\end{array}
\end{equation}
and in the $ \widetilde{\rm \bf NS}$ sector:
\begin{eqnarray*}
\widetilde{\mathcal{F}}_{ \Delta}^{\upodkr\lambda}(q)&=&
q^{{\Delta-\frac{c}{24}+{1\over 16}}} \, \eta(q)^{-\frac32} \, \sqrt{\theta_4(q^{1/2})} \,
 \widetilde{\mathcal{H}}_{ \Delta}^{\upodkr\lambda}(q),
\\[5pt]
\widetilde {\mathcal{H}}_{ \Delta}^{\upodkr\lambda}(q)
    &=&
\sum_{f\in {1\over 2}\mathbb{N}\cup\{0\}} q^{f} \,
\widetilde    H^{\upodkr\lambda, f}_{ \Delta} =
\sum_{f\in {1\over 2}\mathbb{N}\cup\{0\}} (-1)^{2f} \, q^{f} \,
   H^{\upodkr\lambda, f}_{ \Delta}.
\end{eqnarray*}
Coefficients ${H}_{ \Delta}^{\upodkr\lambda,f}$ have
 the same analytic properties as coefficients
 $ F^{\upodkr\lambda, f}_{ \Delta} $ and formula (\ref{residuaN}) yields the  recursive relation:
\begin{eqnarray}
\label{recrel:N}
 H^{\upodkr\lambda, f}_{ \Delta} \;=\;
\delta^{f}_{0} &+&
\sum\limits_{
\begin{array}{c}
\\[-20pt]
\scriptstyle rs\in 2\mathbb N
\\[-7pt]
\scriptstyle 1 < rs \leqslant 2f
\end{array}
}
\frac{A_{rs}P^{rs}_{c}\!\left[^{\ \underline{\hspace*{4pt}}\,\Delta_{\lambda}}_{\Delta_{rs}+\frac{rs}{2}}\right] \,
P^{rs}_{c}\!\left[^{\underline{\hspace*{4pt}}\,\Delta_{\lambda}}_{\ \Delta_{rs}}\right]}{\Delta - \Delta_{rs}}
H^{\upodkr\lambda, f-{rs \over 2}}_{ \Delta_{rs}+\frac{rs}{2}}
 \\
 \nonumber
&+&
 \sum\limits_{
\begin{array}{c}
\\[-20pt]
\scriptstyle rs\in 2\mathbb N +1
\\[-7pt]
\scriptstyle 1 < rs \leqslant 2f
\end{array}
}
\frac{
A_{rs} P^{rs}_{c}\!\left[^{\ \widetilde{\underline{\hspace*{4pt}}\,\Delta_{\lambda}}}_{\Delta_{rs}+\frac{rs}{2}}\right] \,
P^{rs}_{c}\!\left[^{ \underline{\hspace*{4pt}}\,\Delta_{\lambda}}_{\ \Delta_{rs}}\right]
\widetilde{\_\,s_{rs}}        }{\Delta - \Delta_{rs}}
H^{\upodkr\lambda, f-{rs \over 2}}_{ \Delta_{rs}+\frac{rs}{2}}.
\end{eqnarray}

\subsection{${\rm \bf R}$ and $ \widetilde{\rm \bf R}$   sectors}

The large $\beta$ behavior of the blocks with R intermediate states, (\ref{Rasymptot:1}) and  (\ref{Rasymptot:2}),
lead to the following definition of the elliptic blocks:
\begin{eqnarray*}
\mathcal{F}_{ \beta, { \scriptscriptstyle \rm e}}^{\, \lambda(+)}(q)
&=&
{1\over \sqrt{2}}\, q^{{\Delta-\frac{c}{24}}} \, \eta(q)^{-\frac32} \, \sqrt{\theta_2(q^{1/2})}\
\mathcal{H}_{ \beta, { \scriptscriptstyle \rm e}}^{\, \lambda(+)}(q),
\\[4pt]
\mathcal{F}_{ \beta, { \scriptscriptstyle \rm e}}^{\, \lambda(-)}(q)
&=&
\mathcal{H}_{ \beta, { \scriptscriptstyle \rm e}}^{\, \lambda(-)}(q),
\\[4pt]
\mathcal{F}_{ \beta, { \scriptscriptstyle \rm e}}^{\, \ast\lambda(-)}(q)
&=& \mathrm{e}^{i \frac{\pi}{4}} \,\sqrt{2}\, \beta \,
 q^{{\Delta-\frac{c}{24}}} \, \eta(q)^{-\frac32} \, \sqrt{\theta_2(q^{1/2})}\
\mathcal{H}_{ \beta, { \scriptscriptstyle \rm e}}^{\, \ast\lambda(-)}(q).
\end{eqnarray*}
Since $h_{ \beta}^{\, \ast\lambda(+),f}$ vanishes, recursive relation (\ref{ramec}) implies that $\mathcal{F}_{ \beta, { \scriptscriptstyle \rm e}}^{\, \ast\lambda(+)}$
is identically zero
$$
 \mathcal{F}_{ \beta, { \scriptscriptstyle \rm e}}^{\, \ast\lambda(+)}(q)=0.
 $$
 It follows that all 1-point functions of $\tilde \phi_{\lambda,\bar \lambda} $ vanish in the $\widetilde R$ sector:
$$
\langle \tilde \phi_{\lambda,\bar \lambda} \rangle_{\scriptscriptstyle \rm \widetilde R}=0.
$$
 Using recursive relation (\ref{ramec}) and asymptotics (\ref{Rasymptot:2}) one can also show that
\[ \mathcal{H}_{ \beta, { \scriptscriptstyle \rm e}}^{\, \ast\lambda(-)}(q) = \mathcal{H}_{ \beta, { \scriptscriptstyle \rm e}}^{\, \lambda(-)}(q). \]
There are thus only  two independent elliptic blocks
$$
\mathcal{H}_{\beta, { \scriptscriptstyle \rm e}}^{\,  \lambda(\pm)}(q)
   =
\sum_{f\in \mathbb{N}\cup\{0\}} q^{f} \,
{H}_{ \beta, { \scriptscriptstyle \rm e}}^{\lambda(\pm),f}
$$
with coefficients satisfying the recursive relation
\begin{equation}
\label{hramec}
 {H}_{ \beta, { \scriptscriptstyle \rm e}}^{\, \lambda(\pm), f} =
\delta_0^f
 +
\hspace{-10pt}
\sum\limits_{\begin{array}{c}\scriptstyle {1<rs\leqslant 2f}\\[-6pt]
 \scriptstyle {r+s\in 2\mathbb N+1 }
\end{array}}
\hspace{-10pt}
{A_{rs}
P^{rs}_{c}\!\left[_{\pm \beta'_{rs}}^{\;\Delta_{\lambda}}\right]
P^{rs}_{c}\!\left[_{\pm \beta_{rs}}^{\;\Delta_{\lambda}}\right] \over \Delta - \Delta_{rs}}
{H}_{ \beta'_{rs}, { \scriptscriptstyle \rm e}}^{\, \lambda(\pm), f-{rs\over 2}}.
\end{equation}
\section*{Acknowledgments}

The work  was financed by the NCN grant DEC2011/01/B/ST1/01302.
The work of PS was also supported by the Sciex grant 10.054
of the CRUS  and the Kolumb Programme KOL/6/2011-I of FNP.

\appendix
\label{appendixA}
\section{Neveu-Schwarz sector }
In this appendix we shall prove the propositions of Sect. 4.1.

\noindent{\bf Proof of Proposition 1}

Part 1 is  a simple consequence of the NS algebra.
By the same token one has
$$
{\rm deg}_{\Delta} \langle u_i | u_j \rangle \leqslant
{\rm min}
\left\{
{\rm deg}_{\Delta} \langle u_i | u_i\rangle
,
{\rm deg}_{\Delta} \langle u_j | u_j\rangle
\right\}.
$$
Let $u_i = L_{-M}S_{-K} \nu_\Delta $ and $u_j = L_{-N}S_{-L} \nu_\Delta. $
If $\#M +\#K \neq \#N +\#L,$ then part 2 follows from part 1
and the inequality above.

Suppose $\#M +\#K = \#N +\#L$. In this case the
inequality of part 2 is also satisfied. Indeed, calculating the scalar product
$$
\langle L_{-M}S_{-K}\nu_\Delta| L_{-N}S_{-L}\nu_\Delta\rangle =\langle \nu_\Delta| (S_{-K})^\dagger (L_{-M})^\dagger L_{-N}S_{-L} \nu_\Delta\rangle
$$
by the NS algebra rules one can get the maximal degree $\#M +\#K$ if, and only if $L_{-M}S_{-K}\nu_\Delta  =  L_{-N}S_{-L} \nu_\Delta.$

In order to prove part 3 let us observe that by part 1 the product of diagonal terms
$
Q=\prod\limits_{i=1}^{\dim {\cal B}_f} \langle u_i|u_i \rangle
$
is of a maximal degree i.e.
$$
{\rm deg}_{\Delta}\, Q = {\rm deg}_\Delta \det \Big[ \langle u_i|u_j \rangle\Big].
$$
Any other term in the expression for the determinant of $\Big[ \langle u_i|u_j \rangle\Big]$
takes the form
$$
P_\sigma = \prod\limits_{i=1}^{\dim {\cal B}_f} \langle u_i|u_{\sigma(i)} \rangle,
$$
where $\sigma$ is a nontrivial permutation. Let us assume that for all $i$
$$
{\rm deg}_{\Delta}\langle u_i|u_{\sigma(i)} \rangle ={\rm deg}_{\Delta} \langle u_i|u_i \rangle,
$$
hence ${\rm deg}_{\Delta}\,Q={\rm deg}_{\Delta}\, P_\sigma$. On the other hand by Prop.\ 1.2 the equations above imply that
$$
{\rm deg}_{\Delta} \langle u_i|u_{\sigma(i)} \rangle <{\rm deg}_{\Delta}  \langle u_{\sigma(i)}|u_{\sigma(i)} \rangle
$$
for all $i$ and therefore ${\rm deg}_{\Delta}\,Q>{\rm deg}_{\Delta}\, P$ in contradiction with our assumption.
It follows that for an arbitrary nontrivial permutation $\sigma$ there exists at least
one $i$ such that
$$
{\rm deg}_{\Delta}\langle u_i|u_{\sigma(i)} \rangle <{\rm deg}_{\Delta} \langle u_i|u_i \rangle
$$
hence $ {\rm deg}_{\Delta}\,P_\sigma<{\rm deg}_{\Delta}\,Q$.

\rightline{$\Box$}

\noindent{\bf Proof of Proposition 2:}

Let
\(
V(\podkr\nu_{\lambda}) \,:\,{\cal V}_{\Delta}\to {\cal  V}_{\Delta}
\)
be an NS chiral vertex operator with a conformal weight $\Delta_\lambda$. For any $u_i, u_j \in {\cal B}_f$
of the same parity one has
\begin{eqnarray}
\nonumber
\langle u_i|V(\nu_\lambda)|u_j\rangle & = & \rho_{\scriptscriptstyle \rm NN }(u_i, \nu_{\lambda} , u_j)\langle \nu_\Delta|V(\nu_\lambda)|\nu_\Delta\rangle,
\\
\nonumber
\langle u_i|V(*\nu_\lambda)|u_j\rangle & = & \rho^*_{\scriptscriptstyle \rm NN }(u_i, *\nu_{\lambda} , u_j)\langle \nu_\Delta|V(*\nu_\lambda)|\nu_\Delta\rangle.
\end{eqnarray}
If we assume
\[
{\rm deg}_{\Delta}\langle \nu_\Delta|V(\nu_\lambda)|\nu_\Delta\rangle = {\rm deg}_{\Delta}\langle \nu_\Delta|V(*\nu_\lambda)|\nu_\Delta\rangle = 0
\]
then
\begin{eqnarray}
\nonumber
\label{PropA2:add}
\nonumber
{\rm deg}_{\Delta} \,\rho_{\scriptscriptstyle \rm NN }(u_i, \nu_{\lambda} , u_j) & = & {\rm deg}_{\Delta}\langle u_i|V(\nu_\lambda)|u_j\rangle,
\\
\nonumber
{\rm deg}_{\Delta} \,\rho^*_{\scriptscriptstyle \rm NN }(u_i, *\nu_{\lambda} , u_j) & = & {\rm deg}_{\Delta}\langle u_i|V(*\nu_\lambda)|u_j\rangle
\end{eqnarray}
and it is enough to consider the matrix elements $\langle u_i|V(\podkr\nu_\lambda)|u_j\rangle$.
By Proposition 1.2
$$
{\rm deg}_{\Delta} \langle u_i | u_j \rangle <
{\rm max}
\left\{
{\rm deg}_{\Delta} \langle u_i | u_i\rangle ,
{\rm deg}_{\Delta} \langle u_j | u_j\rangle
\right\}.
$$
Suppose
$$
{\rm max}
\left\{
{\rm deg}_{\Delta} \langle u_i | u_i\rangle ,
{\rm deg}_{\Delta} \langle u_j | u_j\rangle
\right\}={\rm deg}_{\Delta} \langle u_j | u_j\rangle.
$$
Calculating matrix elements
$ \langle u_i |V(\podkr\nu_{\lambda})| u_j \rangle$
one can use the Ward identities to move all the NS algebra generators to the right.\\
Let $u_i= L_{-M}S_{-K} \nu_\Delta,\ u_j= L_{-N}S_{-L} \nu_\Delta,$ then
$$
{\rm deg}_{\Delta} \langle u_i | u_i \rangle = \# M +\# K\leqslant {\rm deg}_{\Delta} \langle u_j | u_j\rangle.
$$
The matrix elements
\(
\langle u_i|V(\podkr\nu_\lambda)|u_j\rangle
\)
can  be represented as a linear combination of
$$
 \langle \nu_\Delta |V(\podkr\nu_{\lambda})|(S_{-K})^\dagger (L_{-M})^\dagger  u_j \rangle
$$
and terms of the form
\[
\langle \nu_\Delta |V(\nu_{\lambda})|\widehat{(S_{-K})^\dagger (L_{-M})^\dagger } u_j \rangle,
\hskip 1cm
\langle \nu_\Delta |V(*\nu_{\lambda})|\widehat{(S_{-K})^\dagger (L_{-M})^\dagger } u_j \rangle,
\]
where $\widehat{(S_{-K})^\dagger (L_{-M})^\dagger }$ denotes product $(S_{-K})^\dagger (L_{-M})^\dagger $
with at least one generator removed. The coefficients of this combination are independent of $\Delta.$

Using Ward identities one easily checks that for arbitrary $L_{-P} S_{-Q}\nu_\Delta$
$$
{\rm deg}_{\Delta}\langle \nu_\Delta |V(\podkr\nu_{\lambda})|{L_{-P} S_{-Q}} \nu_\Delta \rangle =0.
$$
This in order imply
$$
{\rm \deg}_\Delta\,\langle \nu_\Delta |V(\podkr\nu_{\lambda})| (S_{-K})^\dagger (L_{-M})^\dagger  u_j \rangle \leqslant \#M + \#K,
$$
and
\begin{eqnarray*}
{\rm deg}_{\Delta}  \langle \nu_\Delta|V(\podkr\nu_{\lambda})|
\widehat{(S_{-K})^\dagger (L_{-M})^\dagger } u_j \rangle & < & \# M+\# K.
\end{eqnarray*}
On the other hand
$$
 \langle \nu_\Delta|V(\podkr\nu_{\lambda})|
(S_{-K})^\dagger (L_{-M})^\dagger  u_j \rangle
=
\langle \nu_\Delta|V(\podkr\nu_{\lambda})|
\nu_\Delta\rangle
 \langle u_i | u_j\rangle
$$
and by assumption
$$
{\rm deg}_{\Delta} \langle \nu_\Delta|V(\podkr\nu_{\lambda})|
(S_{-K})^\dagger (L_{-M})^\dagger  u_j \rangle
<
{\rm deg}_{\Delta}  \langle u_j | u_j\rangle.
$$
Hence
$$
{\rm deg}_{\Delta}\langle u_i|V(\podkr\nu_\lambda)|u_j\rangle <
{\rm deg}_{\Delta}  \langle u_j | u_j\rangle.
$$
If
$
{\rm max}
\left\{
{\rm deg}_{\Delta} \langle u_i | u_i\rangle ,
{\rm deg}_{\Delta} \langle u_j | u_j\rangle
\right\}={\rm deg}_{\Delta} \langle u_i | u_i\rangle
$
 one can repeat the calculations moving all the NS generators to the left.
One thus gets
\[
{\rm deg}_{\Delta}\langle u_i|V(\podkr\nu_\lambda)|u_j\rangle
\; < \;
{\rm max} \{{\rm deg}_{\Delta}  \langle u_i| u_i\rangle,{\rm deg}_{\Delta}  \langle u_j | u_j\rangle\}.
\]

\rightline{$\Box$}

\section{Ramond sector}
We shall first prove  Eq.\ (\ref{equal:dimensionality:of:pm:subspaces}).

Let $q(k,n)$ be the number of partitions of $n$ in $k$ distinct parts.
The corresponding generating function  reads
$$
\sum\limits_{k, n=0}^\infty q(n,k) y^k q^n = \prod\limits_{i=1}^\infty (1+yq^i).
$$
For $y=-1$ it counts the difference between the number of partitions
in an even number of unequal parts and the number of partitions in an odd number of unequal parts.
Hence
$$
\sum\limits_{ n=0}^\infty ({\rm dim}\,{\cal B}_n^+- {\rm dim}\,{\cal B}_n^-)q^n
=
\prod\limits_{i=1}^\infty {1\over 1-q^i}\prod\limits_{i=1}^\infty (1-q^i)=1.
$$
\rightline{$\Box$}

\noindent
{\bf Proof of Proposition 3:}

Part 1 is a simple consequence of the Ramond algebra. Part 2 follows from part 1 and the observation
that maximal possible degree of $ \langle u^+_k | u^-_l \rangle $ is odd while the diagonal elements of the Gram matrix
are of even degrees. The proof of  part 3 parallels the proof of Proposition 1, part 2 while part 4 is proved along the same lines
as Proposition 1, part 3.

\rightline{$\Box$}

\noindent
{\bf Proof of Proposition 4:}

\noindent
We shall prove part 1 using the same method as in the proof of Proposition 2.

In the case of interest Eqs.\ (\ref{dec}) and  (\ref{dec*}) take the form
\begin{eqnarray*}
\langle u_k^+ |V(\nu_{\lambda}) | u_l^+ \rangle &=&
 \rho^{++}_\RRe(u_k^+,\nu_\lambda,u_l^+) \langle w^+_\beta |V(\nu_{\lambda}) | w^+_\beta \rangle
+\rho^{--}_\RRe(u_k^+,\nu_\lambda,u_l^+) \langle w^-_\beta |V(\nu_{\lambda}) | w^-_\beta \rangle
\end{eqnarray*}
and
\begin{eqnarray*}
\langle u_k^+ |V(*\nu_{\lambda}) | u_l^+ \rangle &=&
 \rho^{+-}_\RRo(u_k^+,*\nu_\lambda,u_l^+) \langle w^+_\beta |V(\nu_{\lambda}) | w^-_\beta \rangle
+\rho^{-+}_\RRo(u_k^+,*\nu_\lambda,u_l^+) \langle w^-_\beta |V(\nu_{\lambda}) | w^+_\beta \rangle.
\end{eqnarray*}
By Proposition 3:
$$
{\rm deg}_\beta \langle u^+_k | u^+_l \rangle <
{\rm max} \{
{\rm deg}_\beta \langle u^+_l | u^+_l \rangle
,
{\rm deg}_\beta \langle u^+_k | u^+_k \rangle\} .
$$
Suppose
\begin{equation}
\label{assumption}
{\rm max} \{
{\rm deg}_\beta \langle u^+_l | u^+_l \rangle
,
{\rm deg}_\beta \langle u^+_k | u^+_k \rangle\}
=
{\rm deg}_\beta \langle u^+_k | u^+_k \rangle
\end{equation}
and let $u^+_k= L_{-M}S_{-K} w^+_\beta, u^+_l= L_{-N}S_{-L} w^+_\beta$ (with $\# K,\# L \in 2\mathbb{N}$).
In order to calculate $ \langle u^+_k |V_{\lambda}(\nu)| u^+_l \rangle$
one can use the Ward identities to move all the Ramond algebra generators to the right representing it
as linear combination (with $\beta-$independent coefficients) of
$$
 \langle w^+ |V(\nu_{\lambda})|
(S_{-K})^\dagger (L_{-M})^\dagger  u_l^+ \rangle
$$
and terms of the form
$$
 \langle w^+ |V(\nu_{\lambda})|
\widehat{(S_{-K})^\dagger (L_{-M})^\dagger } u_l^+ \rangle,\;\;\;
 \langle w^+ |V(*\nu_{\lambda})|
\widehat{(S_{-K})^\dagger (L_{-M})^\dagger } u_l^+ \rangle.
$$
For arbitrary $L_{-P} S_{-Q}w^\pm_\beta$ the Ward identities give
$$
{\rm deg}_\beta
\langle w^+ |V(\podkr\nu_{\lambda})|
{L_{-P} S_{-Q}} w^\pm_\beta \rangle \leqslant 1,
$$
what in turn implies
\begin{eqnarray*}
{\rm deg}_\beta  \langle w^+ |V(\podkr\nu_{\lambda})|
\widehat{(S_{-K})^\dagger (L_{-M})^\dagger } u_l^+ \rangle &< & 2 (\# M+\# K) = {\rm deg} \langle u^+_k | u^+_k \rangle.
\end{eqnarray*}
On the other hand
$$
 \langle w^+ |V(\nu_{\lambda})|
(S_{-K})^\dagger (L_{-M})^\dagger  u_l^+ \rangle
=
\langle w^+ |V(\nu_{\lambda})|
w^+ \rangle
 \langle u_k | u_l^+\rangle
$$
and by assumption
$$
{\rm deg} \langle w^+ |V(\nu_{\lambda})|
(S_{-K})^\dagger (L_{-M})^\dagger  u_l^+ \rangle
<
{\rm deg}  \langle u_k^+ | u_k^+\rangle.
$$
If $
{\rm max} \{
{\rm deg}_\beta \langle u^+_l | u^+_l \rangle
,
{\rm deg}_\beta \langle u^+_k | u^+_k \rangle\}
=
{\rm deg}_\beta \langle u^+_l | u^+_l \rangle
$ one follows similar calculations
moving all the Ramond generators to the left.

Taking into account decomposition (\ref{dec}) one gets in particular
\begin{eqnarray*}
{\rm deg} \,
\rho^{++}_\RRe(u^+_k,\nu_\lambda,u^+_l)
<
{\rm max}
\left\{
{\rm deg} \langle u^+_k | u^+_k\rangle,
{\rm deg} \langle u^+_l | u^+_l\rangle
\right\}.
\end{eqnarray*}
Since the terms
$$
\langle w^+ |V(\nu_{\lambda})|
S_{K}^\dagger L_{M}^\dagger  u_l^+ \rangle,\;\;\;
\langle S_{L}^\dagger L_{N}^\dagger  u_k^+ |V(\nu_{\lambda})|
 w^+ \rangle,
$$
do not contribute to $\rho^{--}_\RRe$ one also has the second inequality of part 3:
\begin{eqnarray*}
{\rm deg} \,
\rho^{--}_\RRe(u^+_k,\nu_\lambda,u^+_l)
<
{\rm min}
\left\{
{\rm deg} \langle u^+_k | u^+_k\rangle,
{\rm deg} \langle u^+_l | u^+_l\rangle
\right\}.
\end{eqnarray*}

The matrix element $ \langle u^+_k |V(*\nu_{\lambda})| u^+_l \rangle$ can be analyzed in a similar way.
Suppose that equation  (\ref{assumption}) holds.
As before one has
\begin{eqnarray*}
{\rm deg}  \langle w^+ |V(\podkr\nu_{\lambda})|
\widehat{(S_{-K})^\dagger (L_{-M})^\dagger } u_l^+\rangle &< & 2 (\# M+\# K).
\end{eqnarray*}
On the other hand
\begin{eqnarray*}
 \langle w^+ |V(*\nu_{\lambda})|
(S_{-K})^\dagger (L_{-M})^\dagger  u_l^+ \rangle
&=&
\langle w^+ |V(*\nu_{\lambda})|
w^+ \rangle
 \langle u_k^+ | u_l^+\rangle
 \\
 &=&
  \beta \, {\rm e}^{ i{\pi\over 4}}\left(\langle w^+ |V(\nu_{\lambda}) | w^- \rangle
+ i \langle w^- |V(\nu_{\lambda}) | w^+ \rangle\right) \langle u_k^+ | u_l^+\rangle
\end{eqnarray*}
and by assumption
$$
{\rm deg} \,\beta^{-1}\langle w^+ |V(*\nu_{\lambda})|
S_{K}^\dagger L_{M}^\dagger  u_l^+ \rangle
<
{\rm deg}  \langle u_k^+ | u_k^+\rangle,
$$
hence
\begin{eqnarray*}
{\rm deg} \,\beta^{-1}
\rho^{+-}_\RRo(u^+_k,*\nu_\lambda,u^+_l)
&<& {\rm deg} \langle u^+_k | u^+_k\rangle,
\\
{\rm deg} \,\beta^{-1}
\rho^{-+}_\RRo(u^+_k,*\nu_\lambda,u^+_l)
&<&
 {\rm deg} \langle u^+_k | u^+_k\rangle.
\end{eqnarray*}
The rest of the proof parallels the considerations above.

\rightline{$\Box$}

\section{Some explicite formulae for blocks' coefficients  }

First elliptic block coefficients of the NS sector as defined in  (\ref{elliptic blocks NS}):
{\footnotesize\begin{eqnarray*}
 {H}_{ \Delta}^{\, \lambda, 0} &=& 1,
\qquad  {H}_{ \Delta}^{\, \lambda, \frac12} = \frac{\Delta_\lambda}{2 \Delta} ,
\qquad
{H}_{ \Delta}^{\, \lambda, 1} = \frac{\Delta^2_\lambda}{2 \Delta} ,
\\
{H}_{ \Delta}^{\, \lambda, \frac32} &=&
\frac{\Delta_\lambda \Big(16 b^2 \Delta^2 + (2 + 5 b^2 +   2 b^4) \Delta_\lambda^2 +
   2 \Delta (2 + 2 b^4 +      b^2 (8 + (-8 +\Delta_\lambda) \Delta_\lambda))\Big)}{2 \Delta (1 +
   2 b^2 + 2 \Delta) (2 + b^2 (1 + 2 \Delta))}
\\
{H}_{ \Delta}^{\, \lambda, 2} &=&
\Delta_\lambda \Big( 6 (1+ b^8 ) \Delta_\lambda (1 + 4 \Delta + \Delta_\lambda^2)
   \\&&\hspace{10pt}   + (b^2 +  b^6) \big(-64 \Delta (1 + 2 \Delta) + (3 + 4 \Delta) (17 + 54 \Delta) \Delta_\lambda
- 24 (1 + 4 \Delta) \Delta_\lambda^2 +
        11 (3 + 2 \Delta) \Delta_\lambda^3 \big)
\\&& \hspace{10pt} +
     b^4 \big(-32 \Delta (5 + 4 \Delta (1 + \Delta)) +
        2 (51 + 2 \Delta (101 + 92 \Delta + 48 \Delta^2)) \Delta_\lambda -
        4 (3 + 4 \Delta) (5 + 6 \Delta) \Delta_\lambda^2
\\
&& \hspace{20pt}+ (57 +  2 \Delta (19 + 8 \Delta)) \Delta_\lambda^3 \big) \Big)
 \Big(4 \Delta (1 + 2 b^2 +
     2 \Delta) (2 + b^2 (1 + 2 \Delta)) (3 + 3 b^4 + b^2 (6 + 8 \Delta)) \Big)^{-1}
     \\
 {H}_{ \Delta}^{\ast \lambda, 0} &=& 1,
\qquad  {H}_{ \Delta}^{\ast \lambda, \frac12} = \frac{1- 2 \Delta_\lambda}{4 \Delta} ,
\qquad
{H}_{ \Delta}^{\ast \lambda, 1} = \frac{(1- 2\Delta_\lambda)^2}{8 \Delta} ,
\\
{H}_{ \Delta}^{\ast \lambda, \frac32} &=&
-\Big((-1 + 2 \Delta_\lambda) \big(2 (8 \Delta + (1 - 2 \Delta_\lambda)^2)
+
    2 b^4 (8 \Delta + (1 - 2 \Delta_\lambda)^2)
+
    b^2 (64 \Delta^2 + 5 (1 - 2 \Delta_\lambda)^2
\\&&
 \ \ \ +
       \Delta (34 - 72 \Delta_\lambda + 8 \Delta_\lambda^2))\big)\Big)
\,
\Big(
 16 \Delta (1 + 2 b^2 + 2 \Delta) (2 + b^2 (1 + 2 \Delta)) \Big)^{-1}
\\
 {H}_{ \Delta}^{\ast \lambda, 2}
 &=&
(-1 + 2 \Delta_\lambda) \Big(
 6 (1 + b^8 ) (-1 + 2 \Delta_\lambda) (5 +  16 \Delta - 4 \Delta_\lambda + 4 \Delta_\lambda^2)
\\
&& \hspace{44pt}+ (b^2 + b^6) \Big(32 \Delta^2 (5 + 54 \Delta_\lambda) +
        2 \Delta (-119 + 986 \Delta_\lambda - 516 \Delta_\lambda^2 +
           88 \Delta_\lambda^3)
\\&& \hspace{60pt}
+
        3 (-63 + 202 \Delta_\lambda - 196 \Delta_\lambda^2 +
           88 \Delta_\lambda^3)\Big)
           \\&&
\hspace{44pt}
+     b^4 \big(256 \Delta^3 (1 + 6 \Delta_\lambda) +
        16 \Delta^2 (-17 + 190 \Delta_\lambda - 60 \Delta_\lambda^2 +
           8 \Delta_\lambda^3)
\\ && \hspace{60pt}
+
        2 \Delta (-35 + 1730 \Delta_\lambda - 836 \Delta_\lambda^2 +
     3 (-115 + 386 \Delta_\lambda - 388 \Delta_\lambda^2 +
           152 \Delta_\lambda^3)\big)\Big)
\\
&& \times\Big(64 \Delta (1 + 2 b^2 + 2 \Delta) (2 +
     b^2 (1 + 2 \Delta)) (3 + 3 b^4 + b^2 (6 + 8 \Delta))\Big)^{-1}
\end{eqnarray*}
}
First R blocks coefficients as defined in (\ref{blockCoeffR}) and (\ref{blockCoeffR*}):
{\footnotesize
\begin{eqnarray*}
{F}_{ \beta, { \scriptscriptstyle \rm e}}^{\, \lambda(+), 0}&=&1
\\
{F}_{ \beta, { \scriptscriptstyle \rm e}}^{\, \lambda(+), 1} &=&
\frac{6 (2 + 5 b^2 + 2 b^4) (3 + 4 (-1 + \Delta_\lambda) \Delta_\lambda) +
 64 (3 + 3 b^4 + b^2 (3   -6\Delta_\lambda + 2\Delta_\lambda^2 )) \Delta_\beta +
 512 b^2 \Delta_\beta^2}{(3 + 6 b^2 + 16 \Delta_\beta) (6 + b^2 (3 + 16 \Delta_\beta))}
\\
 {F}_{ \beta, { \scriptscriptstyle \rm e}}^{\, \lambda(+), 2}&=&
4 \Bigg(180 (1+ b^8)\Big(11 + 2 (-1 + \Delta_\lambda) \Delta_\lambda + 16 \Delta_\beta \Big) \Big(3 + 4 (-1 + \Delta_\lambda) \Delta_\lambda + 16 \Delta_\beta \Big)
\\&&
\hspace{10pt}+  b^4 \Big(15 (3855 + 2 \Delta_\lambda (-3077 + 5 \Delta_\lambda (1021 + 4 \Delta_\lambda (-124 + 37 \Delta_\lambda))))
\\&&
\hspace{20pt}+
        32 (6078 + \Delta_\lambda (-8961 + \Delta_\lambda (9083 + 224 (-12 + \Delta_\lambda) \Delta_\lambda))) \Delta_\beta
\\&&\hspace{20pt} +
        512 (779 + \Delta_\lambda (-337 + \Delta_\lambda (353 + 4 (-12 + \Delta_\lambda) \Delta_\lambda))) \Delta_\beta^2
\\&& \hspace{20pt}+
        8192 (14 + \Delta_\lambda (-11 + 5 \Delta_\lambda)) \Delta_\beta^3 + 65536 \Delta_\beta^4\Big)
\\&& \hspace{10pt}+
     12 (b^2+  b^6 ) \Big(16 \Delta_\lambda^4 (59 + 24 \Delta_\beta) - 8 \Delta_\lambda^3 (361 + 496 \Delta_\beta) +
        8 \Delta_\lambda^2 (823 + 128 \Delta_\beta (13 + 7 \Delta_\beta))
\\&&
\hspace{20pt} + (13 + 48 \Delta_\beta) (213 +
           32 \Delta_\beta (7 + 8 \Delta_\beta)) - 32 \Delta_\lambda (136 + 21 \Delta_\beta (17 + 16 \Delta_\beta))\Big)\Bigg)
\\
&&\times\Big(
(3 + 6 b^2 + 16 \Delta_\beta) (11 +
     30 b^2 + 16 \Delta_\beta) (6 + b^2 (3 + 16 \Delta_\beta)) (30 + b^2 (11 + 16 \Delta_\beta))
\Big)^{-1}
\end{eqnarray*}
\begin{eqnarray*}
 {F}_{ \beta, { \scriptscriptstyle \rm e}}^{\, \lambda(-), 0}&=& 1
\\
 {F}_{ \beta, { \scriptscriptstyle \rm e}}^{\, \lambda(-), 1}&=&
\frac{32 (4 (\Delta_\lambda -2) \Delta_\lambda +3) \Delta b^2+6 \left(2 b^4+5 b^2+2\right) \left(4 \Delta_\lambda
   ^2-1\right)}{\left(6 b^2+16 \Delta+3\right) \left((16 \Delta +3) b^2+6\right)}
\\
  {F}_{ \beta, { \scriptscriptstyle \rm e}}^{\, \lambda(-), 2} &=& 4 (-1 +
      2\Delta_\lambda) \\
&&
\Bigg( 180 (1 + b^8) (1 + 2\Delta_\lambda) (1 +  2\Delta_\lambda^2 + 8\Delta)  +
\\
&&
\hspace{6pt}12 (b^2 + b^6) \Big(3 (39 + 88\Delta)+  4 \big(\Delta_\lambda (89 +
               2\Delta_\lambda (-33 + 59\Delta_\lambda))
\\&& \hspace{20pt}
        +
            4\Delta_\lambda (149 +
               2\Delta_\lambda (-47 +
                  6\Delta_\lambda))\Delta +
            64 (-3 + 8\Delta_\lambda)\Delta^2\big)\Big)
\\&&
 \hspace{6pt}
+
      b^4 \Big(4\Delta_\lambda^3 (2775 +
            128\Delta (7 + 2\Delta)) -
         2\Delta_\lambda^2 (4725 +
            2432\Delta (7 + 2\Delta))
\\&& \hspace{20pt} -
         9 (-305 +
            8\Delta (31 +
               32\Delta (1 + 8\Delta))) +
         6\Delta_\lambda (1525 +
            8\Delta (1047 +
               32\Delta (39 +
                  8\Delta)))\Big)
\Bigg)\\
&&
\Big((3 + 6 b^2 +
      16\Delta) (11 + 30 b^2 + 16\Delta) (6 +
      b^2 (3 + 16\Delta)) (30 +
      b^2 (11 + 16\Delta))
\Big)^{-1}
\end{eqnarray*}

\begin{eqnarray*}
{F}_{ \beta, { \scriptscriptstyle \rm e}}^{\, \ast\lambda(-), 0} &=& 2  \beta \mathrm{e}^{i {\pi \over 4}}
\\
{F}_{ \beta, { \scriptscriptstyle \rm e}}^{\, \ast\lambda(-), 1} &=& 8 \beta \mathrm{e}^{i {\pi \over 4}}
\Big( 6 + 12 \Delta_\lambda^2 + 48 \Delta+ 6 b^4 (1 + 2 \Delta_\lambda^2 + 8 \Delta)
\\&&
\hspace{25pt}+
   b^2 \big(15 + 6 (4 - 3 \Delta_\lambda) \Delta_\lambda + 8 \Delta + 32 \Delta_\lambda (2 + \Delta_\lambda) \Delta +
      128 \Delta^2\big)\Big)
\\
&&
\times\Big((3 + 6 b^2 + 16 \Delta) (6 + b^2 (3 + 16 \Delta))\Big)^{-1}
\\
 {F}_{ \beta, { \scriptscriptstyle \rm e}}^{\, \ast\lambda(-), 2} &=& 8 \beta \mathrm{e}^{i {\pi \over 4}}
\Bigg(180 (1 + b^8)
 (1 + 2 \Delta_\lambda^2 +  8 \Delta_\beta) (21 + 4 \Delta_\lambda^2 +  32 \Delta_\beta)
\\&&  \hspace{26pt}+
     b^4 \Big( 15 \big(2387 + 2 \Delta_\lambda (-122 + 5 \Delta_\lambda (699 +  4 \Delta_\lambda (-50 + 37 \Delta_\lambda)))\big)
\\&&  \hspace{40pt}  +
        8 \big(29337 + 4 \Delta_\lambda (-5794 + \Delta_\lambda (7627 +
                 224 (-10 + \Delta_\lambda) \Delta_\lambda))\big)
	 \Delta_\beta
\\&&  \hspace{40pt} +
        512 \big(809 + \Delta_\lambda (-214 + \Delta_\lambda (327 +
                 4 (-10 + \Delta_\lambda) \Delta_\lambda))\big) \Delta_\beta^2
\\&&  \hspace{40pt}
 +     2048 (71 +  20 (-2 + \Delta_\lambda) \Delta_\lambda)
\Delta_\beta^3 + 65536 \Delta_\beta^4 \Big)
\\&& \hspace{26pt}
     +  12 (b^2 + b^6) \Big(1729 -  200 \Delta_\lambda^3 (5 + 16 \Delta_\beta) +
        16 \Delta_\lambda^4 (59 + 24 \Delta_\beta)  
\\ && \hspace{77pt}
       - 2 \Delta_\lambda (61 +  128 \Delta_\beta (21 + 34 \Delta_\beta))+
        8 \Delta_\beta (1677 +
           32 \Delta_\beta (65 + 48 \Delta_\beta))
\\&&  \hspace{77pt} +
        4 \Delta_\lambda^2 (1167 +
           32 \Delta_\beta (87 + 56 \Delta_\beta))\Big) \Bigg)
\\&&\times
\Big(\big(3 +  6 b^2 + 16 \Delta_\beta \big) \big(11 + 30 b^2 +
     16 \Delta_\beta \big) \big(6 + b^2 (3 + 16 \Delta_\beta)\big) \big(30 +
     b^2 (11 + 16 \Delta_\beta)\big)\Big)^{-1}
\end{eqnarray*}

}

\end{document}